 \pdfoutput=1

\documentclass[11pt]{article}

\usepackage[final]{acl}

\usepackage{times}
\usepackage{latexsym}
\usepackage{booktabs}
\usepackage{tabularx}
\usepackage[T1]{fontenc}
\usepackage{amsmath} 
\usepackage{multirow}
\usepackage{todonotes}
\usepackage{tcolorbox}
\usepackage{rotating} 
\usepackage{multicol}
\usepackage[commandnameprefix=always]{changes}
\usepackage{wrapfig}

\usepackage[utf8]{inputenc}

\usepackage{microtype}

\usepackage{inconsolata}

\usepackage{graphicx}
\usepackage{pifont}  
\usepackage{threeparttable}
\usepackage{threeparttable, xcolor, colortbl} 

\title{
\begin{minipage}{.1\textwidth}
\centering
\includegraphics[width=\linewidth]{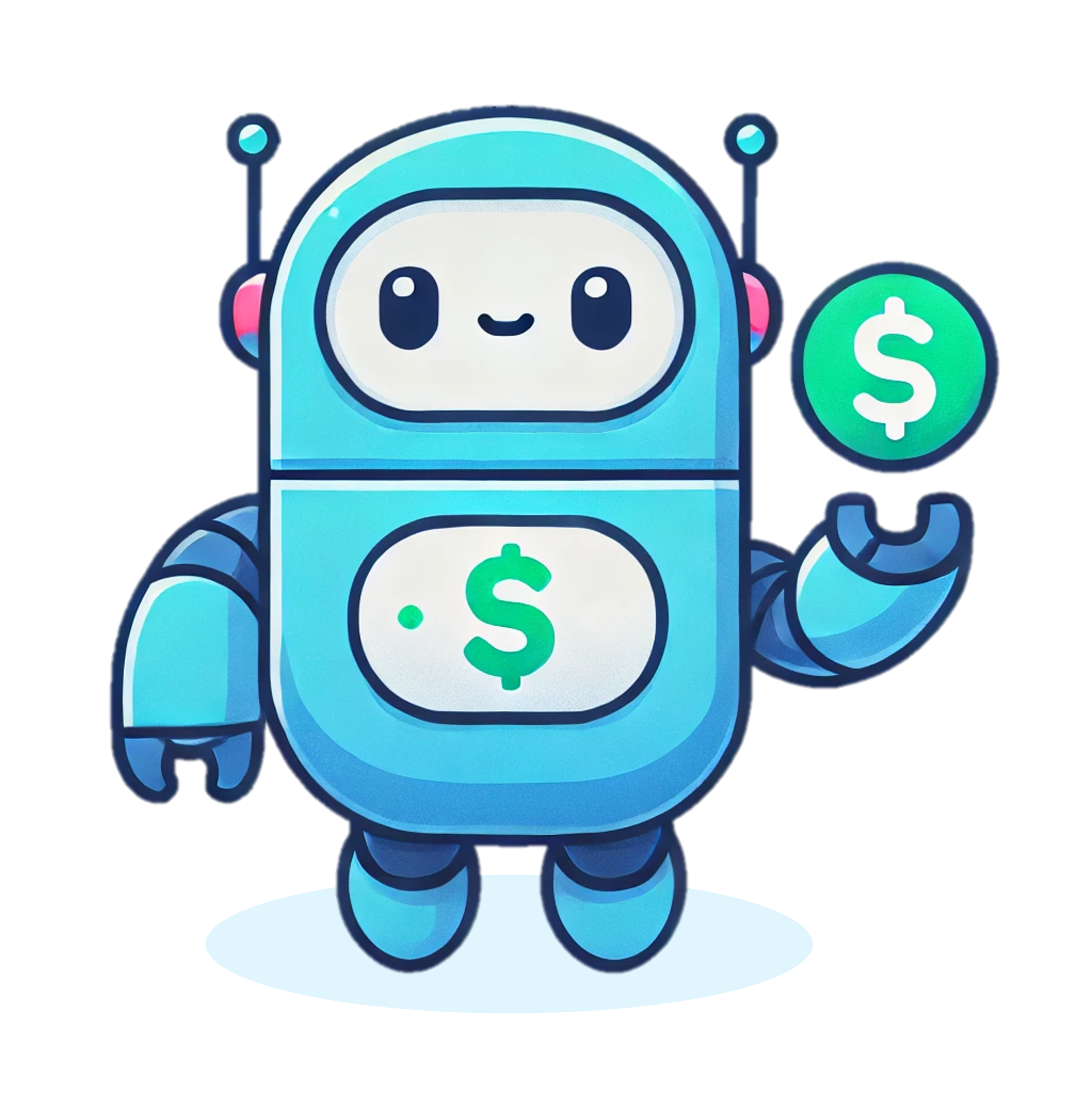} 
\end{minipage}
{\color{blue}UCFE}: A {\color{blue}U}ser-{\color{blue}C}entric {\color{blue}F}inancial {\color{blue}E}xpertise Benchmark for Large Language Models
}

\author{
  \textbf{Yuzhe Yang\textsuperscript{1}\thanks{Equal contribution: \texttt{yuzheyang@link.cuhk.edu.cn}, \texttt{yf\_zhang@smail.nju.edu.cn}, \texttt{huyan@cuhk.edu.cn}}},
  \textbf{Yifei Zhang\textsuperscript{2}\footnotemark[1]},
  \textbf{Yan Hu\textsuperscript{1}\footnotemark[1]},\\
  \textbf{Yilin Guo\textsuperscript{1}},
  \textbf{Ruoli Gan\textsuperscript{1}},
  \textbf{Yueru He\textsuperscript{3}},
  \textbf{Mingcong Lei\textsuperscript{1}},
  \textbf{Xiao Zhang\textsuperscript{3}},
  \textbf{Haining Wang\textsuperscript{2}},\\ 
  \textbf{Qianqian Xie\textsuperscript{3}}\footnotemark[2],
  \textbf{Jimin Huang\textsuperscript{3}},
  \textbf{Honghai Yu\textsuperscript{2}\footnotemark[2]},
  \textbf{Benyou Wang\textsuperscript{1}\thanks{Corresponding authors: \texttt{qianqian.xie@thefin.ai}, \texttt{hhyu@nju.edu.cn}, \texttt{benyouwang@cuhk.edu.cn}}} \\
  \textsuperscript{1}The Chinese University of Hong Kong, Shenzhen,
  \textsuperscript{2}Nanjing University, 
  \textsuperscript{3}The Fin AI \\
  \url{https://github.com/TobyYang7/UCFE-Benchmark}
}

\begin{document}
\maketitle
\begin{abstract}
This paper introduces the UCFE: User-Centric Financial Expertise benchmark, an innovative framework designed to evaluate the ability of large language models (LLMs) to handle complex real-world financial tasks. UCFE benchmark adopts a hybrid approach that combines human expert evaluations with dynamic, task-specific interactions to simulate the complexities of evolving financial scenarios. Firstly, we conducted a user study involving 804 participants, collecting their feedback on financial tasks. Secondly, based on this feedback, we created our dataset that encompasses a wide range of user intents and interactions. This dataset serves as the foundation for benchmarking 11 LLMs services using the LLM-as-Judge methodology. Our results show a significant alignment between benchmark scores and human preferences, with a Pearson correlation coefficient of 0.78, confirming the effectiveness of the UCFE dataset and our evaluation approach. \textbf{UCFE benchmark} not only reveals the potential of LLMs in the financial domain but also provides a robust framework for assessing their performance and user satisfaction.

\end{abstract}

\section{Introduction}\label{intro}
Recent advances in large language models (LLMs) have expanded their potential applications in finance \cite{wu2023bloomberggpt, huang2023finbert, kim2024financial}. Finance professionals are increasingly using LLMs to solve specialized financial tasks \cite{li2023cfgpt, xie2024openfinllmsopenmultimodallarge, yang2023fingptopensourcefinanciallarge, zhang2023xuanyuan20largechinese}, including explorations into LLM-powered financial agents \cite{li2024econagent,yang2025twinmarket}. Financial tasks are inherently complex, involving specialized context, financial terminologies, legal intricacies, and dynamic markets that involve information with high noise-to-signal ratio \cite{pagano1993financial,mullainathan2017machine,li2018more}, which adds significant challenges for LLMs to address. Accurate analysis of financial information is crucial, as even minor ignorance in a signal or market information can lead to substantial financial losses \cite{tversky1981framing,thaler2008mental,mohamed2024size}.


To be effective, LLMs need to swiftly adapt to fiscal policy changes, market fluctuation, extreme events, and global factors, identifying key signals within real-time data to manage volatility and mitigate risks \cite{gueta2024can,yadav2024generative}. Financial markets can react instantly to news, making it crucial for LLMs to process information in near real-time by rapidly consolidating unstructured, real-time data from multiple sources \cite{nguyen2023generative,tong2024ploutos}. Despite LLMs' improving accuracy on tasks like sentiment analysis, market prediction, and risk assessment \cite{wimmer2023leveraging,lopez2023can,rizinski2024comparative}, these models still face significant limitations, such as their reliance on static datasets and challenges in handling real-time data, which hinders their real-world applicability in dynamic financial contexts. Moreover, the evolving nature of financial regulation adds another layer of complexity, requiring LLMs to continuously update their knowledge to remain compliant and useful \cite{yao2024survey,he2024s}. These limitations highlight the need for a more dynamic evaluation framework that assesses LLMs' performance under real-time, evolving financial conditions, ensuring they can handle not only static tasks but also the unpredictable nature of real-world financial environments.



\begin{figure*}[t]
    \centering
    \includegraphics[width=1\linewidth]{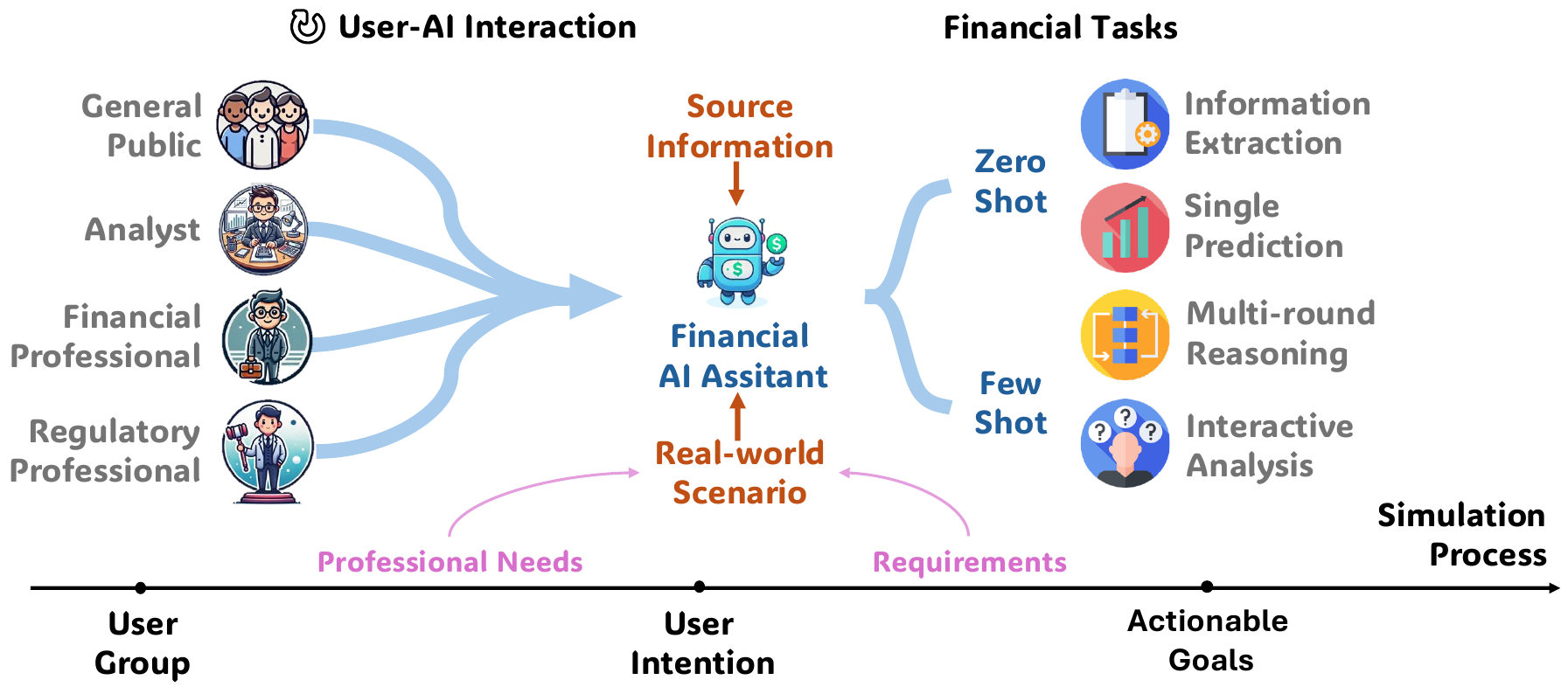}
    \caption{Overview framework of the \textbf{UCFE Benchmark}.}
    \label{fig:intro}
\end{figure*}

To address these challenges in financial domain, we propose a novel framework, the \textbf{{\color{blue}U}ser-{\color{blue}C}entric {\color{blue}F}inancial {\color{blue}E}xpertise Benchmark}, designed to evaluate the ability of LLMs to handle financial tasks in real-world scenarios. Figure~\ref{fig:intro} provides an overview of the framework. The \textbf{UCFE benchmark} has the following key features:
 
\textbf{User-Centric Design:}
Based on preliminary surveys and research, we categorized the target user group into four distinct types: analysts, financial professionals, regulatory professionals, and the general public. Using questionnaires in Appendix~\ref{questionnaire}, we gathered insights into the primary needs and practical applications of each group. This allowed us to refine the user categories for more targeted evaluations, in which LLMs are prompted to simulate roles representative of each group. We developed 17 task types tailored to these user profiles, encompassing 330 data points that include multi-round dialogues in both zero-shot and few-shot settings. More details of the dataset will be explained in Section~\ref{sec:dataset}.


\textbf{Dynamic Interactions:}
In the few-shot tasks, each user group follows a task-oriented approach, where an initial action goal is defined. Users articulate their professional needs and specific task requirements through successive interactions. To simulate real-world user scenarios, we employ authentic datasets that closely mirror actual financial scenarios. This dynamic interaction setup ensures that LLMs are not only responding to isolated queries but are also engaging in an evolving dialogue, adjusting their responses based on the ongoing professional needs expressed by the user. This method provides a more accurate reflection of how LLMs would perform in practical, task-specific financial contexts. 


In summary, this work makes the following key contributions: (1) We propose a new framework that combines human expert judgments with LLMs to assess their ability to handle increasingly complex financial tasks. (2) By leveraging dynamic, user-centric interactions, this work probes the boundaries of LLM capabilities by examining how well LLMs adapt to evolving professional needs and increasingly complex task requirements, which provides a deeper understanding of their potential and limitations in addressing real-world financial scenarios.

\section{Related Work}



\subsection{Financial Benchmark} \label{sec:fin_ben}

FLARE \cite{xie2023pixiu} evaluates models on five financial tasks \footnote{sentiment analysis, news headline classification, named entity recognition, question answering, and stock movement prediction}. Existing benchmarks \cite{zhang2023finevalchinesefinancialdomain,li2023chatgpt,yuan2024r} primarily use multiple-choice questions to assess models' domain knowledge, with questions sourced from real-world financial documents and publicly available financial reports or websites, covering a wide range of topics such as finance, economy, accounting, and certification. MMMU and MMMU-PRO \cite{yue2024mmmu,yue2024mmmu2} extend beyond traditional financial NLP tasks by incorporating multimodal inputs to better evaluate models in more complex financial tasks. Although these benchmarks have advanced the evaluation of financial language models, they predominantly consider structured NLP tasks with deterministic answers and rely heavily on multiple-choice questions or tasks with specific answers, such as sentiment analysis and named entity recognition. This limits their ability to assess generative capabilities, which are essential for simulating real-world financial applications \cite{krause2023large,koa2024learning}.


\subsection{User-Centric Framework}

The implementation of user-centric models involves integrating users into core business processes to harness their creative potential, which has been successfully demonstrated by companies like LEGO, IBM, and Coloplast~\cite{hienerth2011exploring,kwon2021enterprise}. EUCA framework, a practical prototyping toolkit designed to make AI systems explainable to non-technical end-users, provides twelve end-user-friendly explanatory forms that do not require technical knowledge to bridge the gap between technical creators and non-technical users \cite{jin2021euca}. In the financial domain, user-centric explainability is also crucial in algorithmic decision-making systems like robo-advisors~\cite{naveed2022explainable,roveda2023human,pisoni2023responsible}. Research has highlighted the importance of providing transparent and comprehensible explanations to users, which indicated that user trust and confidence in financial applications are positively correlated with the transparency and comprehensibility of the explanations provided~\cite{how2020artificial,deo2021user,xu2024ai}.






\section{Background}

Recent advancements in LLMs have demonstrated significant potential in addressing complex financial tasks. Numerous organizations are now actively training their own LLMs, aiming to enhance their performance by incorporating extensive domain-specific knowledge. For instance, FinGPT \cite{yang2023fingptopensourcefinanciallarge}, which applies supervised fine-tuning to the LLaMA model, has shown notable improvements in financial tasks. Through continued pretraining, models like FinLLaMA \cite{xie2024openfinllmsopenmultimodallarge} have further advanced LLM performance across various metrics. These developments highlight the growing demand and potential for LLMs in the financial domain, both in academia and industry.

From a technical standpoint, LLMs have undoubtedly reduced costs and improved efficiency by quickly processing vast amounts of financial text data with commendable performance. However, the challenges of developing real-world financial applications extend beyond technical issues. These challenges include business requirements, industry-specific barriers, data privacy concerns, accountability, and ethical considerations \cite{nie2024survey, yao2024survey}, along with a gap in understanding between LLMs, functioning as AI assistants, and the specific needs of financial experts. 

As discussed in Section \ref{sec:fin_ben}, existing benchmarks largely focus on technical metrics such as accuracy and efficiency, often ignoring these broader challenges. By emphasizing only technical aspects, such benchmarks fail to address the real-world complexities of financial applications, where business rules, legal frameworks, and human judgment play crucial roles. This makes non-technical aspects, particularly human-AI interaction in finance, comparatively under-explored. Human-AI interaction is critical in financial settings, as it affects decision-making, user trust, and the effective integration of AI systems into the financial workflow. Without considering these factors, current benchmarks offer an incomplete evaluation, limiting the practical relevance of LLMs for real-world financial applications. There is a pressing need for evaluation frameworks that not only assess technical performance but also account for the nuanced interplay between AI systems and financial professionals in complex environments.

In addition to these challenges, the rise of FinTech companies such as Robinhood \footnote{\url{https://robinhood.com/us/en/}} has spurred increasing public interest in finance and trading. A growing number of individuals, many without formal financial education, are seeking accessible ways to manage their finances and participate in the market. For these users, LLMs have become a significant source of financial learning and advice. However, the accuracy, flexibility, and contextual understanding of LLMs are crucial to meet this emerging demand. Models must not only provide correct and relevant information but also adapt to the diverse financial needs of the general public, from basic education to advanced financial management strategies.

In this context, the \textbf{UCFE benchmark} will mainly focus on evaluating the interaction between LLMs and humans, as the improvements in user experience often have greater practical significance than gains in task-specific metrics. By introducing this new framework, we aim to offer deeper insights into the future development of financial LLMs, aligning model performance more closely with human preferences across multiple dimensions. This approach is intended to provide a more holistic understanding of how LLMs can better serve real-world financial applications, ultimately leading to more user-centric AI solutions.

\section{User-Centric Financial Expertise Dataset} \label{sec:dataset}

\subsection{User Preference Alignment} \label{sec:expert1}

To align our dataset more closely with real-world financial tasks and user needs, we conducted a survey to gather insights into how users engage with financial scenarios. Participants completed a questionnaire designed to capture key aspects of their interactions with financial tasks, focusing on their roles, levels of experience, and the types of tasks they typically perform. The survey included questions about participants' familiarity with financial analysis, preferred sources of information, and their engagement preferences regarding financial tasks.

Feedback was solicited from participants across three main areas:
\begin{itemize}
    \item \textbf{Participant Demographics}: Information on the participants' backgrounds and expertise.
    \item \textbf{Detailed Interaction with Financial Tasks}: Assessment of participants' experiences and interactions with specific financial tasks.
    \item \textbf{Financial Scenario Coverage}: Evaluation of how well the tasks reflected real-world financial scenarios.
\end{itemize}
The full questionnaire can be found in Appendix~\ref{questionnaire}.

\subsection{Dataset Construction}

\begin{table}[t]
\resizebox{1\linewidth}{!}{
\begin{tabular}{lrrr}
\toprule
 & \textbf{User}  & \textbf{Familiarity}  & \textbf{Importance}  \\    \midrule
Total             & 804    &  458    & 660 \\ \midrule
Student (Finance-related)   & 167    & 148     &155   \\
Financial Professional  & 83     & 83      & 83  \\
Regulatory Professional    & 51       & 47      & 50  \\ \midrule
General Public     & 136     &  49    & 82  \\
Non-Finance Professional & 87     & 37      & 70  \\
Student (Non-finance)  & 208     & 79      & 163  \\ 
Other         & 72      & 15      & 57  \\ 
\bottomrule
\end{tabular}
}
\caption{The user survey outcomes. Familiarity indicates the results of Question 5, where people choose ``they have encountered multi-round financial tasks''. Importance indicates the results of Question 6 where people choose ``they think multi-round financial tasks are important''. }
\label{table:questionnare_res}
\vspace{-2ex}
\end{table}

Based on the results of the survey shown in Table \ref{table:questionnare_res}, we recognized the necessity of constructing a multi-round finance dialogue benchmark that serves both the finance-related and non-finance groups. The survey revealed diverse user intentions and varying levels of financial expertise, underscoring the need for a benchmark that can accommodate a broad spectrum of scenarios. By catering to these different groups, we aimed to capture a comprehensive range of dialogue interactions, from complex financial analysis to simpler, more general financial inquiries, ensuring the dataset reflects real-world variations in user needs and knowledge levels.

To establish the multi-round dialogue benchmark, we meticulously selected sources that encompass authoritative financial reports, regulatory documents, and accessible online resources based on the survey results shown in Appendix \ref{res_sur}. This selection process was designed to ensure that the dataset meets both the technical demands of financial professionals and the practical needs of general users. By synthesizing insights from diverse user experiences and expert evaluations, we aimed at creating a dataset that facilitates effective multi-round interactions, ultimately enhancing the user experience in financial analysis.

\subsection{Tasks} \label{task}

\begin{figure}[t]
        \centering
        \includegraphics[width=0.7\linewidth]{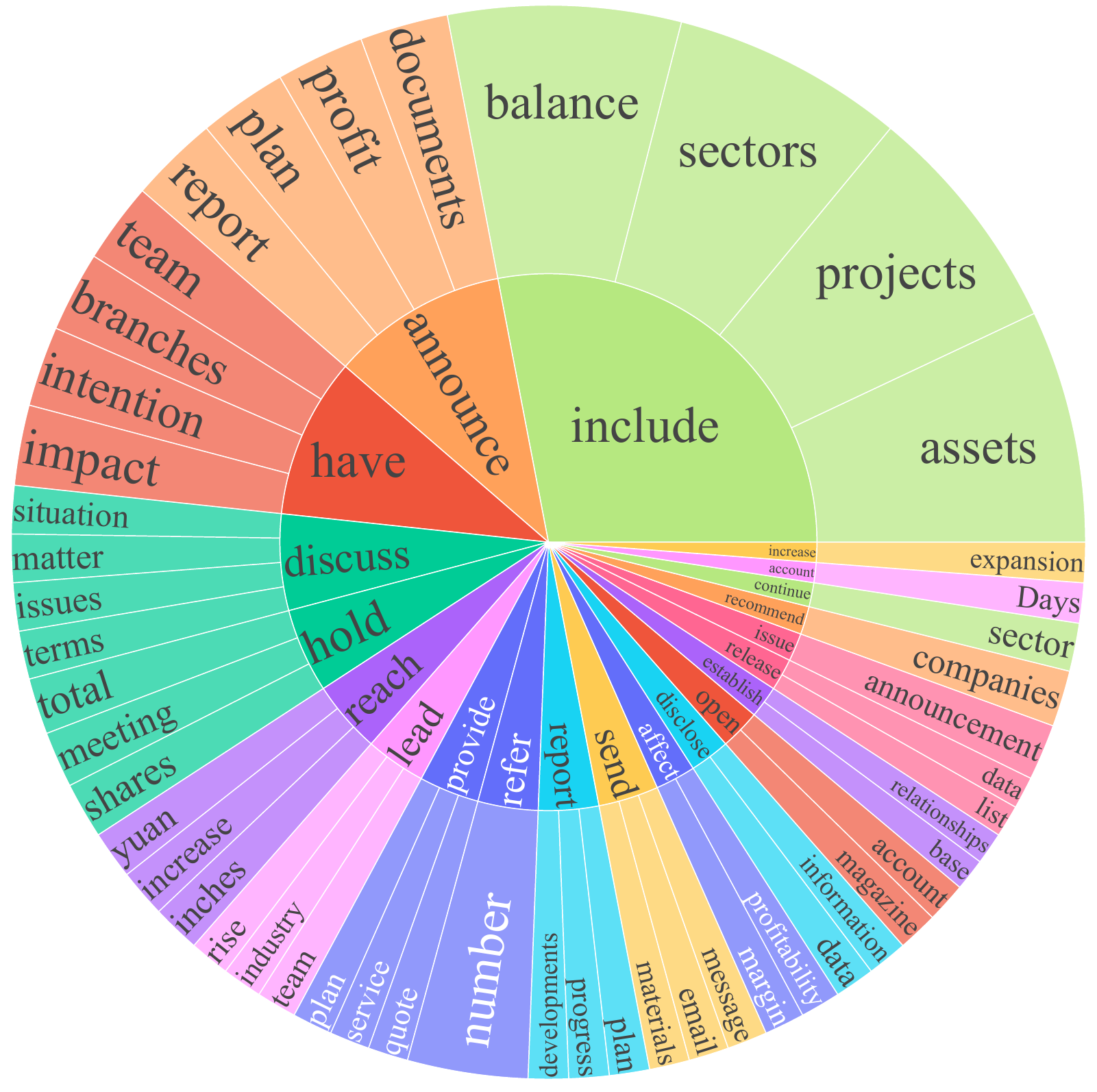}
        \caption{The visualization displays the top 25 most common root verbs (inner circle) and their top 4 associated direct noun objects (outer circle) extracted from the provided texts. 
        }
        \label{fig:sunburst}
\end{figure}

\begin{figure}[t]
        \centering
        \includegraphics[width=1\linewidth]{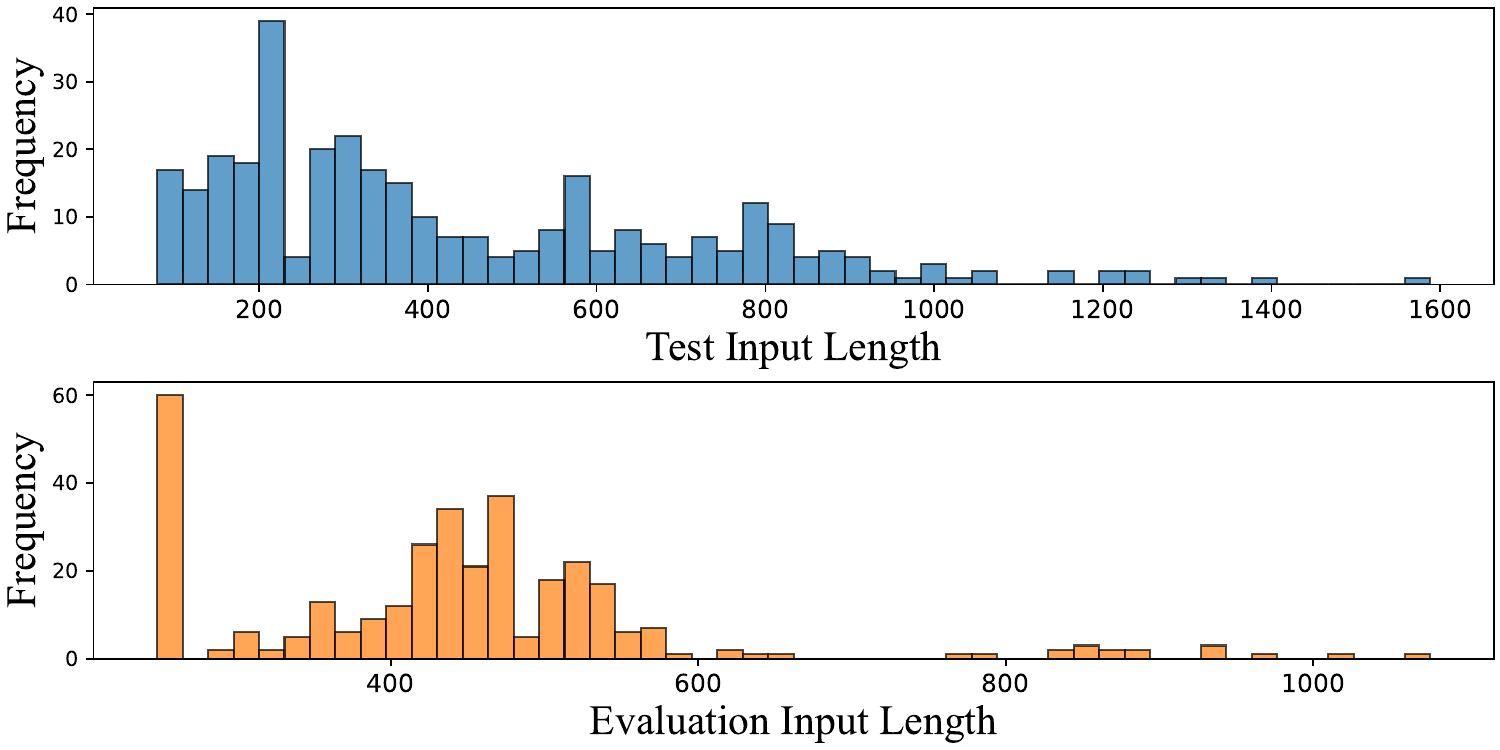}
        \caption{Distribution of test and evaluation input lengths for the datasets.}
        \label{fig:length}
\end{figure}

Table \ref{tab:source_info} and  \ref{table:dataset} provide the statistical breakdown of the \textbf{UCFE benchmark}, with all data sourced from the previous user survey targeting various user groups. The \textbf{UCFE benchmark} encompasses both few-shot and zero-shot tasks, with a total of 17 distinct tasks covering a broad range of financial scenarios. These tasks are specifically designed to reflect practical financial needs, including but not limited to market information summarization, asset valuation, and regulatory compliance assessments. The multi-turn nature of these tasks emphasizes dynamic user interaction and adaptive decision-making.

Figure \ref{fig:sunburst} presents the visualization of the 25 most common root verbs (inner circle) and their top 4 associated direct noun objects (outer circle), providing insights into the types of financial interactions covered by the dataset. The diversity of verb-noun pairs highlights the wide range of financial operations and decision-making processes represented, ensuring the benchmark tasks reflect the complex and varied language used in financial contexts. 

In addition, Figure \ref{fig:length} shows the distribution of input lengths for both test and evaluation queries, revealing significant variance in task complexity. Shorter queries require concise outputs, while longer inputs demand deeper comprehension and detailed responses. This variance challenges models not only to generalize across different task types but also to adapt their performance based on the complexity of the input, making it essential for evaluating LLMs' scalability and versatility in real-world financial tasks.

\begin{table*}[t]
\centering
\scriptsize
\begin{threeparttable}
\resizebox{\textwidth}{!}{
\begin{tabular}{@{}l l l l@{}}
\toprule
\textbf{Category} 
& \textbf{Task} 
& \textbf{Source}
& \textbf{Target User Group} \\ 
\midrule
\multirow{12}{*}{\textbf{Few-shot}} 
& Analyst Simulation & TCL Annual Report \& Analyst Report & Senior Analyst \\
& Asset Valuation Reporting & EastMoney & Analyst  \\ 
& Company Evaluation Reporting & Analyst Report & Analyst \\ 
& Corporate Operation Analysis & Analyst Report & Analyst  \\ 
& Credit Risk Evaluation & GPT-4 Generated & Analyst   \\
& Financial Knowledge Consulting & Investopedia\tnote{1} & General Public \& Financial Professional  \\ 
& Financial Regulation Consulting & Securities Law\tnote{2} & General Public \& Financial Professional \& Regulatory Professional   \\ 
& Industry Report Summarization & EastMoney & General Public \& Financial Professional  \\ 
& Insider Trading Detection & Securities Regulatory Commission\tnote{3} &  Regulatory Professional   \\
& Investment Strategy Evaluation & Seeking Alpha\tnote{4} & Analyst  \\
& Investment Strategy Optimization & Financestrategists\tnote{5} & Analyst  \\
& Newshare Evaluation Reporting & Stock.us\tnote{6} & Analyst  \\
& Prospectus Risk Summarization & Prospectus \& Inquiry Letter\tnote{7} &   General Public \& Financial Professional  \\ 
\midrule
\multirow{4}{*}{\textbf{Zero-shot}} 
& Stock Price Prediction & A-stock Statistics & General Public \& Financial Professional \\ 
& Negative Information Detection & EastMoney  & General Public \& Financial Professional  \\ 
& Financial Indicator Calculation & CPA \& CFA & General Public \& Financial Professional  \\ 
& Financial Text Summarization & News Headlines & General Public \& Financial Professional   \\ 
\bottomrule
\end{tabular}
}
\begin{tablenotes}
    \scriptsize
    \item[1] \url{https://www.investopedia.com/financial-term-dictionary-4769738}
    \item[2] \url{https://www.gov.cn/xinwen/2019-12/29/content_5464866.htm}
    \item[3] \url{http://www.csrc.gov.cn/csrc/c101953/zfxxgk_zdgk.shtml}
    \item[4] \url{https://seekingalpha.com/article/4500869-portfolio-performance-evaluation-metrics}
    \item[5] \url{https://www.financestrategists.com/wealth-management/investment-management/portfolio-performance-evaluation/}
    \item[6] \url{https://stock.us/cn/stock/sz/001279}
    \item[7] \url{https://www.sse.com.cn/disclosure/credibility/supervision/inquiries/}
\end{tablenotes}
\end{threeparttable}
\caption{Overview of \textbf{UCFE benchmark} tasks, including task categories, sources, and target user groups.}
\label{tab:source_info}
\end{table*}

\begin{table}[t]
\centering
\resizebox{1\linewidth}{!}{ 
\begin{tabular}{lrr}
\toprule
\textbf{Task Type} & \textbf{Number of Tasks} & \textbf{Number of Questions} \\ 
\midrule
Zero-shot Tasks  & 4  & 80  \\ 
Few-shot Tasks   & 13 & 250 \\ 
\midrule
\textbf{Total}   & 17 & 330 \\ 
\bottomrule
\end{tabular}
}
\caption{Summary of Task Types and Corresponding Number of Questions in the \textbf{UCFE benchmark}. \textbf{Note that all tasks have 20 questions except that ``Analyst Simulation'' has only 10 questions.}}
\label{table:dataset}
\vspace{-2ex}
\end{table}

\section{UCFE Benchmark}

\begin{figure*}[t]
    \centering
    \includegraphics[width=1\linewidth]{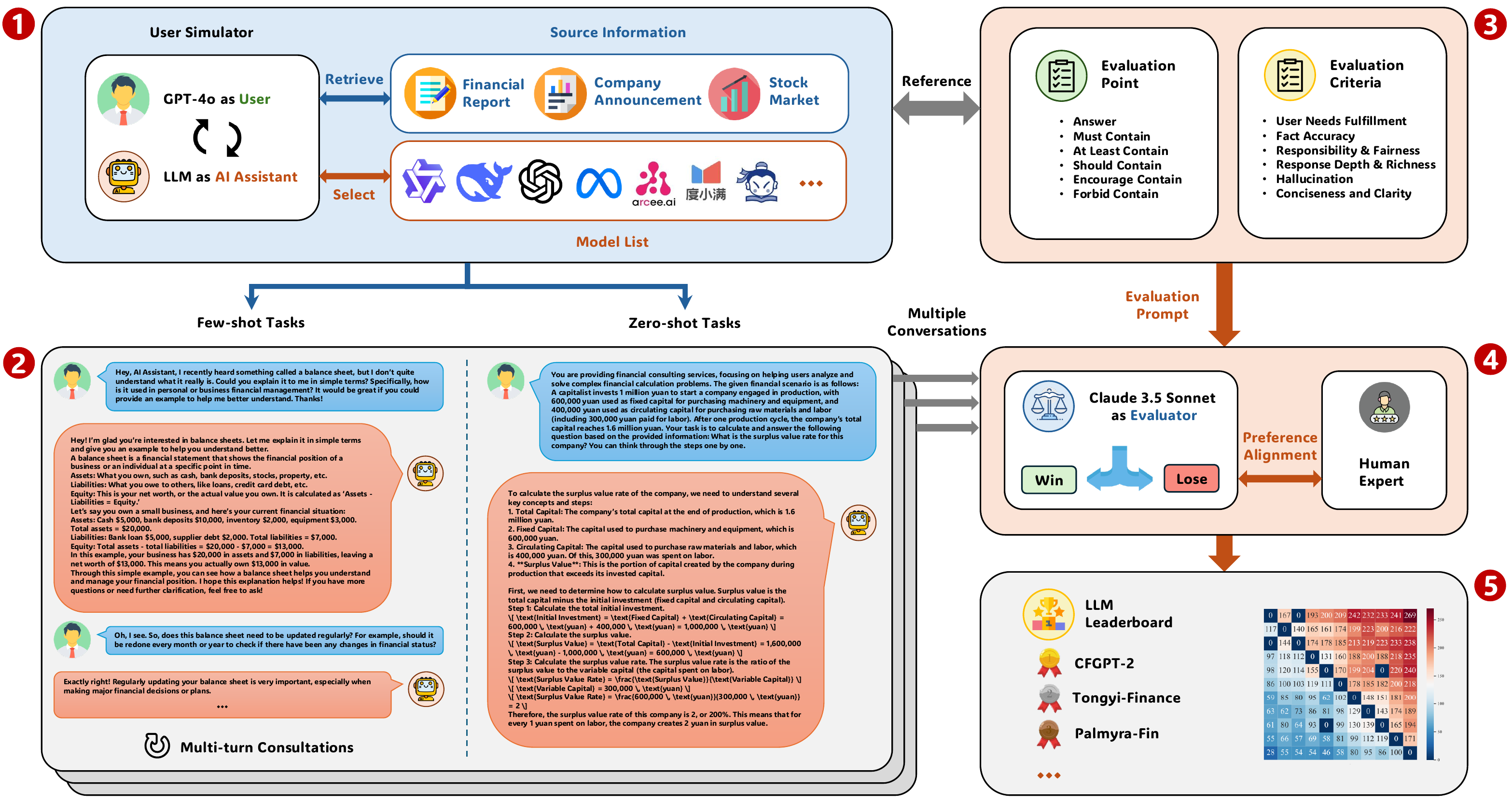}
    \caption{The evaluation pipeline of the \textbf{UCFE Benchmark} involves the following steps: \ding{172} selecting the model and task, \ding{173} generating dialogues between the user and AI assistant via a user simulator, \ding{174} creating evaluation prompts based on source information to assess model performance, \ding{175} pairwise comparison of dialogue outputs by evaluators, aligned with human expert judgments, and \ding{176} computing Elo scores based on win-loss outcomes.}
    \label{fig:framework}
\end{figure*}

In this section, we provide an overview of the technical details and evaluation pipeline of the \textbf{UCFE benchmark}. As shown in Figure~\ref{fig:framework}, the evaluation starts by selecting finance-specific tasks (introduced in Section~\ref{task}), where the model acts as an AI assistant. GPT-4o\footnote{\url{https://openai.com/index/gpt-4o-system-card/}} simulates user interactions, generating dialogue data based on realistic behavior. Using LLMs to simulate user roles is common in recent research \cite{inaba2024largelanguagemodelsused}. To minimize model bias, we established evaluation criteria (detailed in Section \ref{sec:setting}). Model outputs are then compared in pairs, with Claude-3.5-Sonnet\footnote{\url{https://www.anthropic.com/news/claude-3-5-sonnet}} as the evaluator, following the common practice LLm-as-judge framework for evaluation \cite{liu2023gevalnlgevaluationusing}. Each model’s performance is measured using Elo scores, which offer a relative comparison of capabilities. Finally, results are compared against human expert preferences to ensure evaluation robustness.

\subsection{Evaluation Method} \label{sec:method}

We use the Elo rating system for model evaluation, which is well-suited for comparing multiple models. This system is widely applied in competitive environments, such as match result prediction in sports like association football \cite{hvattum2010using,chiang2024chatbot}. Its key advantages are:

\begin{itemize}
    \item \textbf{Dynamic Adjustments}: Elo ratings are continuously updated based on relative model performance, making it ideal for frequent comparisons.
    \item \textbf{Scalability \& Efficiency}: New models can be added without retesting all previous ones, saving time and API costs.
\end{itemize}

Each model starts with an Elo rating of $1000$, which is updated after every comparison task. For each task, dialogues generated by the target model and the base model are compared using specific prompts. A Claude-based model evaluates the comparison as a win, loss, or tie, and the Elo ratings are updated using the formula:

\[
R' = R + K \times (S - E)
\]

where \( R' \) is the updated rating, \( R \) is the current rating, \( S \) is the result (1 for a win, 0.5 for a tie, and 0 for a loss), and \( E \) is the expected result, computed as:

\[
E = \frac{1}{{1 + 10^{\frac{(R_o - R)}{S}}}}
\]

Here, \( R_o \) is the opponent's rating, \( S \) is set to 400, and \( K \) is 4. These parameters control the magnitude of rating updates. This process repeats for each task, and the final Elo ratings reflect the models' comparative performance across all tasks.

\subsection{Experimental Settings} \label{sec:setting}

In the experiments, GPT-4o is used as the user simulator to generate queries and simulate real-world conversations. Claude-3.5-Sonnet serves as the evaluator to compare model responses, ensuring a clear separation between testing and evaluation to minimize bias.

For dialogue simulations, the temperature is set to 0.5 with no token limit. We tested financial-specific LLMs (7B to 70B parameters) along with their backbone models and included general-purpose models like GPT-4o and GPT-4o-mini, accessed via APIs. Table \ref{tab:models} lists all models used.

To mitigate positional bias in LLM evaluations \cite{li2023splitmergealigningposition}, we shuffled the input order during dialogue comparisons. To further minimize evaluator bias, such as misinformation or cognitive bias \cite{talboy2023challengingappearancemachineintelligence}, we designed the evaluation prompts based on two key criteria: 

\begin{itemize}
    \item \textbf{Source Information Content}: Categorized into \texttt{Answer}, \texttt{Must Contain}, \texttt{At Least Contain}, \texttt{Should Contain}, \texttt{Encourage Contain}, and \texttt{Forbid Contain}, guiding LLMs to make accurate content-based choices.
    \item \textbf{Evaluation Standards}: Focused on \texttt{User Needs Fulfillment}, \texttt{Fact Accuracy}, \texttt{Responsibility \& Fairness}, \texttt{Response Depth \& Richness}, \texttt{Hallucination}, and \texttt{Conciseness \& Clarity}, ensuring a thorough assessment.
\end{itemize}

The full evaluation prompt is available in Appendix~\ref{prompt}.

\begin{table}[t]
\centering
\small
\renewcommand{\arraystretch}{1} 
\resizebox{0.5\textwidth}{!}{
\begin{threeparttable} 
\begin{tabular}{@{}l l l@{}}
\toprule
\textbf{Model}
&\textbf{Type}
\\ 
\midrule
\textbf{CFGPT2-7B}\tnote{1} \cite{li2023cfgpt}
& \textbf{Financial}
\\ 
GPT-4o
& General
\\ 
GPT-4o-mini
& General
\\ 
InternLM2.5-7B-Chat \cite{cai2024internlm2}
& General
\\ 
Llama-3.1-70B-Instruct \cite{llama3modelcard}
& General
\\ 
Llama-3.1-8B-Instruct 
& General
\\ 
\textbf{Llama3-XuanYuan3-70B-Chat} \cite{zhang2023xuanyuan20largechinese}
& \textbf{Financial}
\\
\textbf{Palmyra-Fin-70B-32k} \cite{Palmyra-Fin-70B-32k}
& \textbf{Financial}
\\ 
Qwen2.5-14B-Instruct \cite{qwen2.5}
& General
\\ 
Qwen2.5-7B-Instruct 
& General
\\ 
\textbf{Tongyi-Finance-14B-Chat}\tnote{2}
& \textbf{Financial}
\\ 
\bottomrule
\end{tabular}
\begin{tablenotes}
    \scriptsize
    \item[1] The backbone model of CFGPT2-7B is InternLM2-7B.
    \item[2] The backbone model of Tongyi-Finance-14B-Chat is Qwen-14B.
\end{tablenotes}
\end{threeparttable}
}
\caption{Models evaluated in \textbf{UCFE benchmark}.}
\label{tab:models}
\end{table}

\subsection{Overall Results}

Table~\ref{tab:elo_res} presents a comprehensive overview of model performance across the 17 distinct financial tasks within the \textbf{UCFE benchmark}. A key finding is the \textbf{consistent outperformance }of financially-specialized LLMs (Tongyi-Finance-14B-Chat, CFGPT2-7B, and Palmyra-Fin-70B-32k) compared to original backbone models ( Qwen series, InternLM series and Llama series). This performance gap is not uniform across all tasks, highlighting the varied strengths and weaknesses of different model architectures and training strategies when applied to the complexities of the financial domain. While general-purpose models demonstrate a baseline level of competence, their performance often lags significantly on tasks requiring in-depth financial knowledge, precise terminology, and adherence to regulatory constraints.

\begin{figure}[t]
    \centering
    \includegraphics[width=1\linewidth]{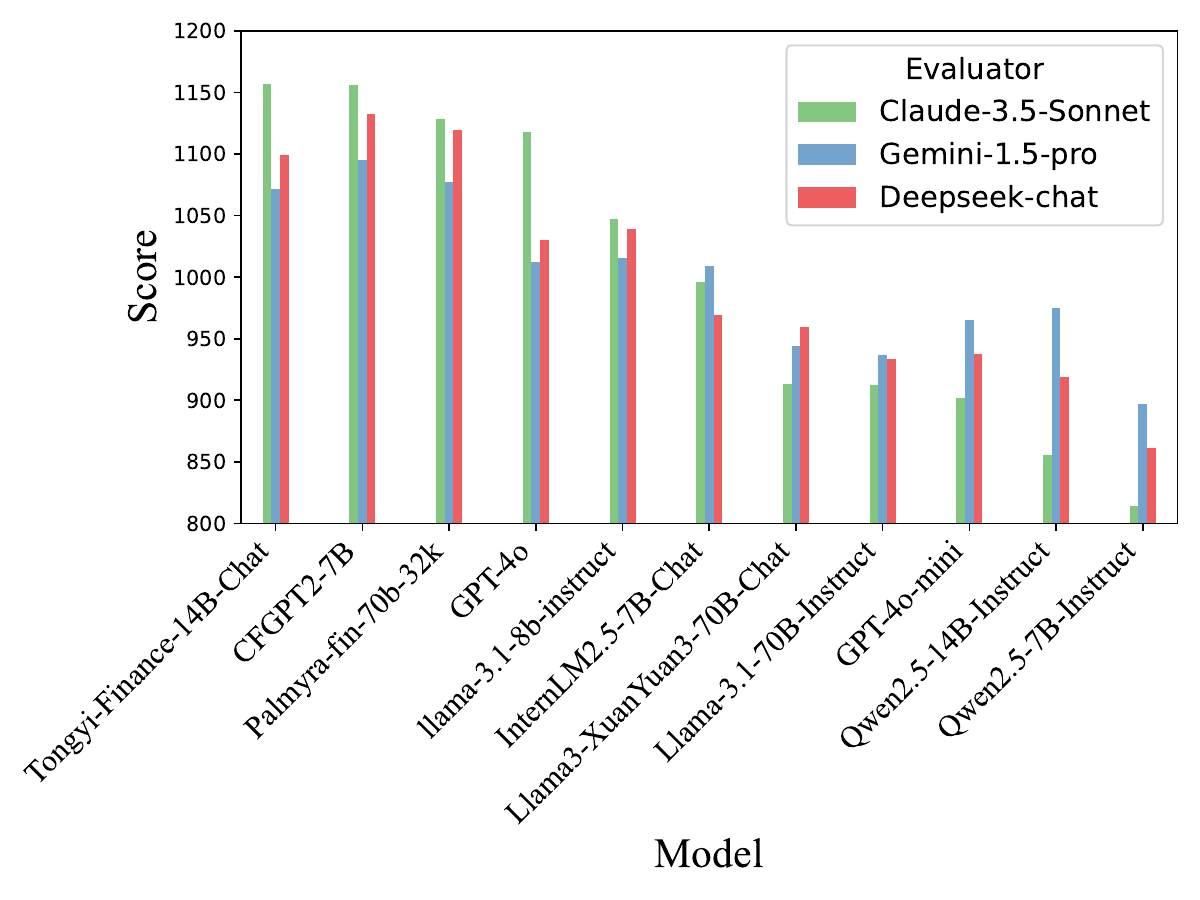}
    \caption{Comparison of model performance on \textbf{UCFE benchmark} across three evaluators.}
    \label{fig:radar}
\end{figure}

To ensure the robustness of our evaluation and mitigate potential biases inherent in using a single LLM as a judge (Claude-3.5-Sonnet), we conducted supplementary evaluations using two additional, independent LLM evaluators: Gemini-1.5-pro\footnote{\url{https://deepmind.google/technologies/gemini/pro/}} and Deepseek-chat\footnote{\url{https://www.deepseek.com/}}. As illustrated in Figure~\ref{fig:radar}, the final Elo scores across the target models show a high degree of consistency across all three evaluators, indicating that the relative rankings of the target models are largely insensitive to the choice of evaluator, suggesting that our findings are not an artifact of any particular evaluator's biases or limitations. 

\begin{figure}[t]
    \centering
    \includegraphics[width=1\linewidth]{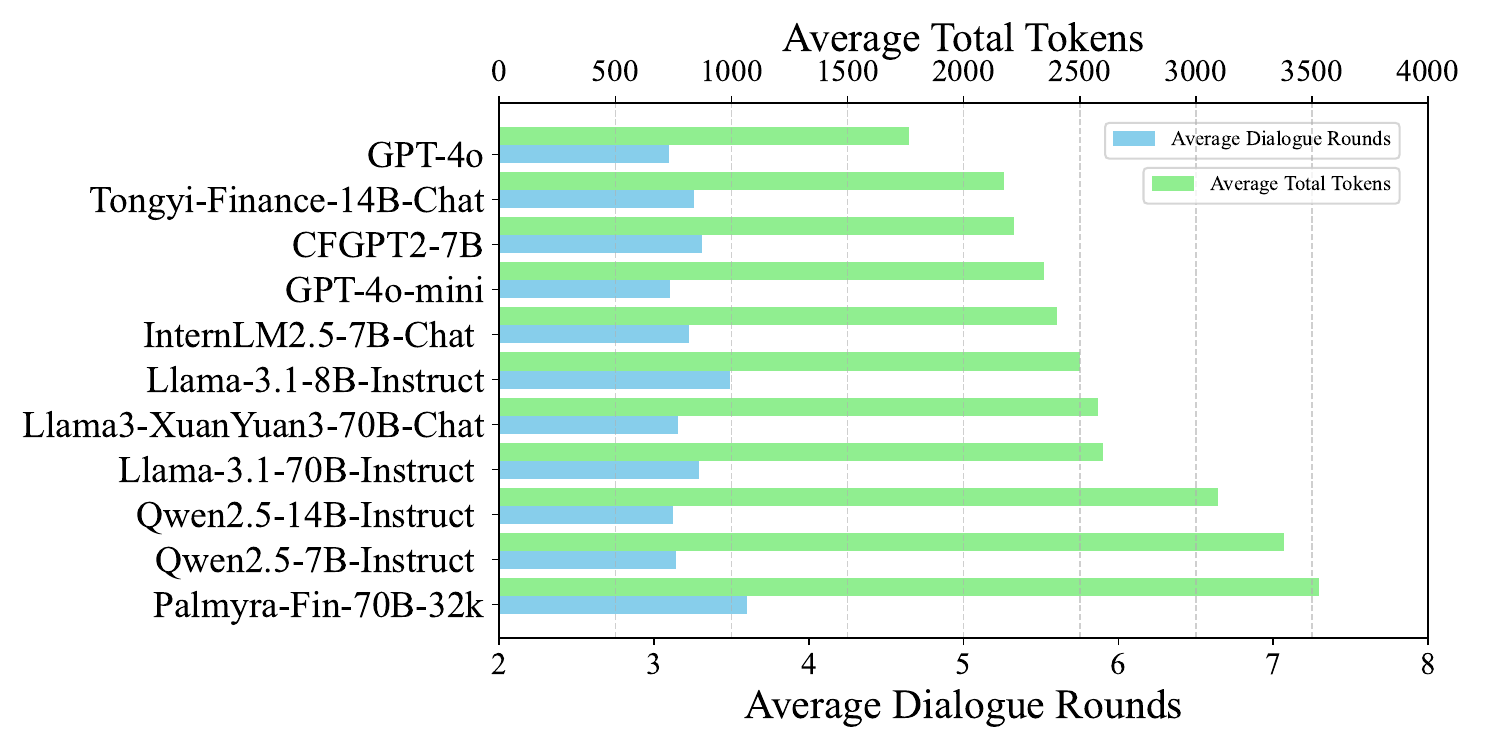}
    \caption{Comparison of average dialogue rounds and total tokens across different models in few shot tasks.}
        \label{fig:token}
\end{figure}

Additionally, to examine potential length bias~\cite{wei2024systematic}, we analyzed dialogue lengths and turn counts across models (Figure~\ref{fig:token}). The analysis shows no significant correlation, indicating no observable bias.

\begin{table}[t]
\centering
\small
\renewcommand{\arraystretch}{1} 
\resizebox{0.5\textwidth}{!}{
\begin{threeparttable} 
\begin{tabular}{@{}l c c c c@{}}
\toprule
\textbf{Model}
& \textbf{Overall}
& \textbf{Zero Shot}
& \textbf{Few Shot}
& \textbf{Win Counts} \\
\midrule
\textbf{Tongyi-Finance-14B-Chat} & \colorbox{red!10}{\textbf{1156.99}} & 1007.52 & \colorbox{red!10}{\textbf{1171.27}} & 3614 \\
\textbf{CFGPT2-7B} & \colorbox{blue!10}{\textbf{1155.75}} & \colorbox{red!10}{\textbf{1125.33}} & \colorbox{blue!10}{\textbf{1157.93}} & \colorbox{red!10}{\textbf{3972}} \\
\textbf{Palmyra-Fin-70B-32k} & 1128.25 & 1028.18 & 1143.66 & \colorbox{blue!10}{\textbf{3634}} \\
GPT-4o & 1117.68 & 979.85 & 1120.89 & 3040 \\
Llama-3.1-8B-Instruct & 1046.87 & \colorbox{blue!10}{\textbf{1062.18}} & 1051.32 & 3294 \\
Internlm2.5-7b-chat & 995.85 & 1009.78 & 1000.52 & 2964 \\
\textbf{Llama3-XuanYuan3-70B-Chat} & 913.48 & 934.51 & 911.59 & 2050 \\
Llama-3.1-70B-Instruct & 912.26 & 986.77 & 906.80 & 2196 \\
GPT-4o-mini & 901.75 & 943.81 & 908.92 & 2326 \\
Qwen2.5-14B-Instruct & 855.82 & 974.27 & 840.05 & 1774 \\
Qwen2.5-7B-Instruct & 814.48 & 946.45 & 786.28 & 1312 \\
\bottomrule
\end{tabular}
\end{threeparttable}
}
\caption{Model results in the \textbf{UCFE benchmark}. \colorbox{red!10}{Red} highlights the highest value, while \colorbox{blue!10}{Blue} represents the second-highest value.}
\label{tab:elo_res}
\end{table}


\subsection{Human Preference Alignment} \label{sec:expert2}

\begin{figure}[t]
    \centering
    \includegraphics[width=1\linewidth]{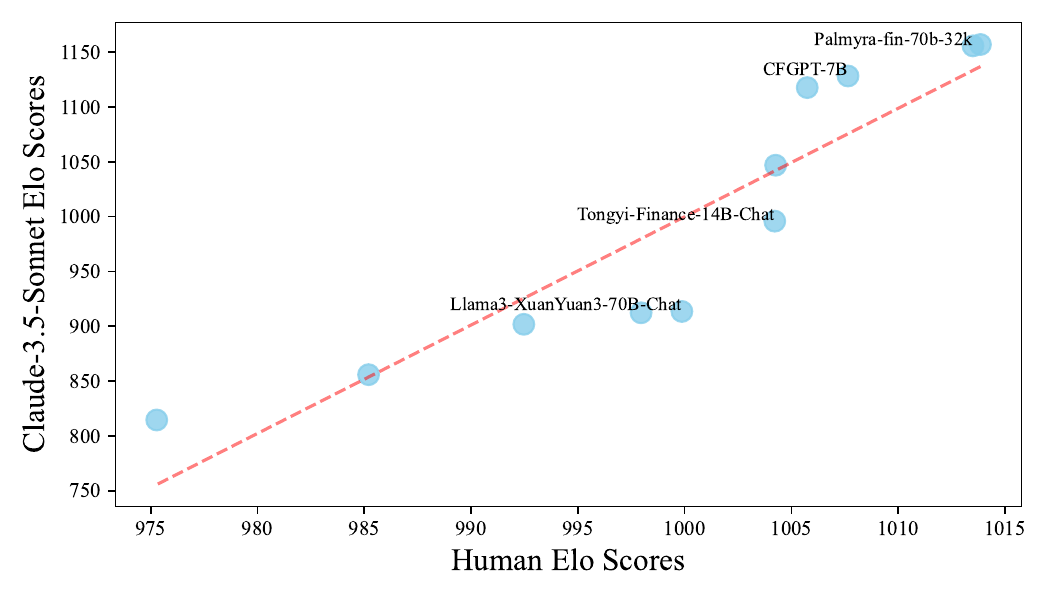}
    \caption{Correlation between human Elo scores and Claude-3.5-Sonnet Elo scores.}
    \label{fig:human}
\end{figure}

To further validate that our results aligned with actual preferences of users, we also conducted a human preference evaluation involving 15 financial professionals and students, with each assigned 10 pre-existing results to conduct the human preference alignment phase. After manually reassessing the outputs, we updated the Elo scores and compared them with our model's predictions. To quantify the similarity between the model's results and human evaluations, we used the Pearson correlation coefficient:

$$
r = \frac{\sum_{i=1}^n (x_i - \bar{x})(y_i - \bar{y})}{\sqrt{\sum_{i=1}^n (x_i - \bar{x})^2 \sum_{i=1}^n (y_i - \bar{y})^2}}
$$

where $x_i$ and $y_i$ represent the Elo scores from participants and model predictions, respectively. The analysis revealed a clear positive correlation, with a calculated Pearson correlation of  \textbf{$r = 0.78$}, as shown in Figure \ref{fig:human}, indicating that the model's performance aligns well with human preferences.

\subsection{Case Study}

\begin{figure}[t]
    \centering
    \includegraphics[width=1\linewidth]{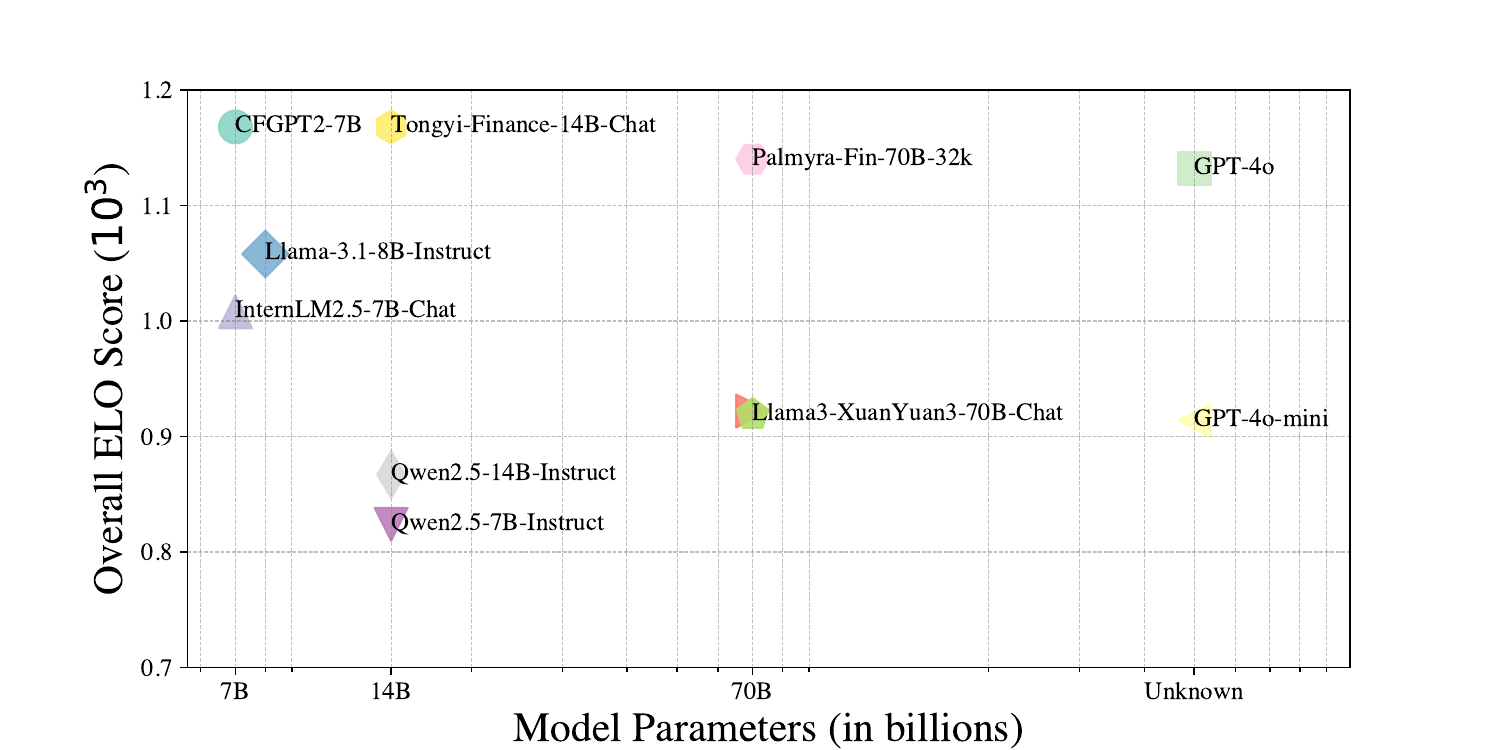}
    \caption{Overall Elo scores of various models plotted against model parameters (in billions).}
        \label{fig:size}
\end{figure}

\begin{figure}[t]
    \centering
    \begin{tcolorbox}[colback=gray!10, colframe=black, width=0.5\textwidth]
    \scriptsize
    \textcolor[rgb]{0.0, 0.5, 0.0}{Here is a summary of the financial text in a single sentence:} \\
    The US Treasury Department \textcolor{red}{\sout{will impose sanctions}} \textcolor[rgb]{0.0, 0.5, 0.0}{has imposed sanctions} on four \textcolor{red}{\sout{Ukrainian government officials}} \textcolor[rgb]{0.0, 0.5, 0.0}{current and former Ukrainian government officials} for \textcolor{red}{\sout{who assisted}} \textcolor[rgb]{0.0, 0.5, 0.0}{their involvement} in a \textcolor[rgb]{0.0, 0.5, 0.0}{Russian} disinformation campaign aimed at undermining \textcolor{red}{\sout{Ukraine's regime and justifying an invasion.}} \textcolor[rgb]{0.0, 0.5, 0.0}{the Ukrainian regime.} \\
    \textcolor[rgb]{0.0, 0.5, 0.0}{Note: I would like to point out that the provided text is not a financial text but rather a political/news article. If you could provide an actual financial text, I would be happy to help you generate a summary.}
    \end{tcolorbox}
    \caption{Comparison between Llama3.1-8B-Instruct and Llama3.1-70B-Instruct models, \textcolor[rgb]{0.0, 0.5, 0.0}{green} highlighting the changes in Llama3.1-70B-Instruct.}
    \label{fig:llama_comparison}
\end{figure}

Previous research has demonstrated the significant influence of scaling laws on model performance \cite{kaplan2020scalinglawsneurallanguage, ruan2024observationalscalinglawspredictability}. As shown in Figure \ref{fig:size}, our results also show a similar trend. LLMs with larger parameters generally outperform smaller ones within the same series, also LLMs from the same backbone model have better results after being trained on financial corpus. However, Llama3.1 appears to be an outlier in this pattern. As shown in Figure \ref{fig:llama_comparison}, we illustrate a result of the \textit{Summarization} task that highlights this phenomenon. In many of the data points, we observe that the 70B model tends to generate longer, more verbose outputs compared to smaller models. This aligns with the conclusions of \cite{chiang2024overreasoningredundantcalculationlarge}, where larger models are prone to over-reasoning, generating lengthy and unnecessary responses to questions. In our evaluation framework, where clear scoring criteria have been established, these redundant outputs significantly lower the model's performance.

\section{Conclusion}

In this paper, we introduced the \textbf{UCFE Benchmark}, a framework designed to evaluate user-AI interactions in the financial domain using the LLM-as-Judge methodology. This framework enables direct comparisons of model performance with human expert preferences while addressing potential biases. Our findings demonstrate that LLMs trained on domain-specific financial texts show notable improvements in understanding complex financial concepts and accurately interpreting user intent. Notably, mid-sized models (7B to 14B parameters) performed particularly well, striking an effective balance between computational efficiency and domain-specific expertise without the excessive overhead of larger models. These results emphasize the importance of optimizing LLMs not only for performance but also for resource efficiency, making them more viable for real-world financial applications. Additionally, the user-centric design of our benchmark highlights the critical role of aligning AI systems with diverse user needs, ensuring that LLMs deliver practical, contextually relevant solutions in finance. This approach lays the foundation for more reliable and scalable AI-driven innovations in the financial industry.

\section*{Limitation}

The limitations of our work can be summarized as follows:

\begin{itemize}
    \item \textbf{Coverage of Financial Tasks:} The financial domain encompasses a wide range of complex tasks and scenarios, from regulatory compliance to dynamic market analysis. While the \textbf{UCFE Benchmark} includes several representative tasks, the diversity and volume of data points may not be sufficient to fully capture all real-world financial applications. This limitation restricts the benchmark's ability to comprehensively assess LLM performance across the entire spectrum of financial use cases.
    
    \item \textbf{Human Preference Bias:} The evaluation framework relies on human preferences to assess model performance, which introduces potential biases. Given the limited number of evaluators and the relatively narrow range of professional backgrounds represented, the results may not fully reflect the diverse needs and preferences of the broader financial community. Individual biases and subjective judgments could influence the evaluation, potentially skewing the assessment of LLM effectiveness in real-world financial tasks.
    
    \item \textbf{Use of Historical Data:} The benchmark relies primarily on historical financial data for task evaluation. While this data is useful for assessing LLM performance in past scenarios, it may not fully capture the evolving and real-time nature of financial markets. This reliance on historical data limits the ability to evaluate how well LLMs can adapt to unforeseen events or respond to rapidly changing market conditions.
\end{itemize}

\section*{Ethical Statements}

We do not see our work to have possible harmful
outcomes. We follow the ACL ethical guidelines
when conducting the research in this paper.

\section*{Acknowledgement}

This work was supported by the Major Program of the National Fund of Philosophy and Social Science of China (No. 19ZDA105), the Shenzhen Science and Technology Program (JCYJ20220818103001002), the Shenzhen Doctoral Startup Funding (RCBS20221008093330065), the Tianyuan Fund for Mathematics of the National Natural Science Foundation of China (NSFC) (12326608), the Shenzhen Key Laboratory of Cross-Modal Cognitive Computing (Grant No. ZDSYS20230626091302006), and the Shenzhen Stability Science Program 2023.


\newpage
\bibliography{custom}

\newpage
\appendix

\section*{Appendix}
\counterwithin{figure}{section}
\counterwithin{table}{section} 
\counterwithin{equation}{section} 
\renewcommand\thefigure{\Alph{section}-\arabic{figure}} 
\renewcommand\thetable{\Alph{section}-\arabic{table}} 
\section{Questionnaire} \label{questionnaire}

The questionnaire of our survey is shown in Figure \ref{fig:Questionnaire}.

\begin{figure*}[htbp]
\centering
\begin{tcolorbox}[title = {Questionnaire}, colframe=gray, colback=gray!10, coltitle=white, colbacktitle=gray!50!black]
\footnotesize

\textbf{We are conducting a study to gather insights on how users engage with financial tasks in real-world scenarios. Your participation will help us improve the design of user-centric and multi-round financial analysis tasks. The survey will take approximately 10 minutes.}

\textbf{\\Section 1: Participant Demographics}

\textbf{\\1. What is your current role or profession?}

A. General Public (No professional experience in finance)

B. Student (Finance-related major)

C. Student (Non-finance major)

D. Finance Professional (e.g., Analyst, Banker, Consultant)

E. Non-Finance Professional (e.g., Engineer, Teacher, etc.)

F. Regulatory Professional (e.g., Securities Regulator, Compliance Officer)

G. Other (please specify)

\textbf{\\2. How familiar are you with financial analysis tasks (e.g., stock price prediction, credit risk evaluation, etc.)?}

A. Not familiar~~~~~B. Somewhat familiar~~~~~C. Very familiar

\textbf{\\3. What is your primary source of financial information?}

A. Company reports (e.g., annual reports, prospectuses)

B. Financial news outlets (e.g., Bloomberg, Reuters)

C. Online financial services (e.g., Yahoo Finance, Eastmoney)

D. Financial consultancies or analysts

E. Other (please specify)

\textbf{\\Section 2: Interaction with Financial Tasks}

\textbf{\\1. How often do you perform financial analysis tasks at work or in your personal life?}

A. Daily~~~~~B. Weekly~~~~~C. Monthly~~~~~D. Rarely

\textbf{\\2. Have you engaged in financial tasks that involve multi-round analysis (i.e., where multiple steps or iterations are required)?}

A. Yes~~~~~B. No~~~~~C. Not sure

\textbf{\\3. Do you think it is necessary to study multi-round financial tasks, both academically and in the finance industry?}

A. Yes~~~~~B. No~~~~~C. Not sure

\textbf{\\4. When working on financial tasks, do you prefer receiving predefined options (e.g., multiple-choice) or generating your own answers (e.g., writing reports or summaries)?}

A. Predefined options (e.g., multiple-choice)

B. Generating answers (e.g., writing reports, creating strategies)

C. A mix of both

\textbf{\\Section 3: Scenario Coverage}

\textbf{\\1. Which financial tasks have you encountered in your work or studies?} (Open-ended)

\textbf{\\2. Do you find it useful to simulate real-world financial scenarios (e.g., stock market predictionsrisk assessments) when completing tasks?}

A. Yes, it helps to improve my analysis skills

B. Somewhat, but real-world scenarios can be complex

C. No, I prefer hypothetical tasks

\textbf{\\3. Where do you come from?} (Open-ended)

\textbf{\\Note: We collect this questionnaire solely for academic purposes, and your personal information will not be used for commercial purposes.}

\end{tcolorbox}
\caption{The questionnaire of our survey}
\label{fig:Questionnaire}
\end{figure*}

\section{UCFE Dataset Information} \label{dataset_info}

\begin{figure}[t]
    \centering
    \includegraphics[width=0.8\linewidth]{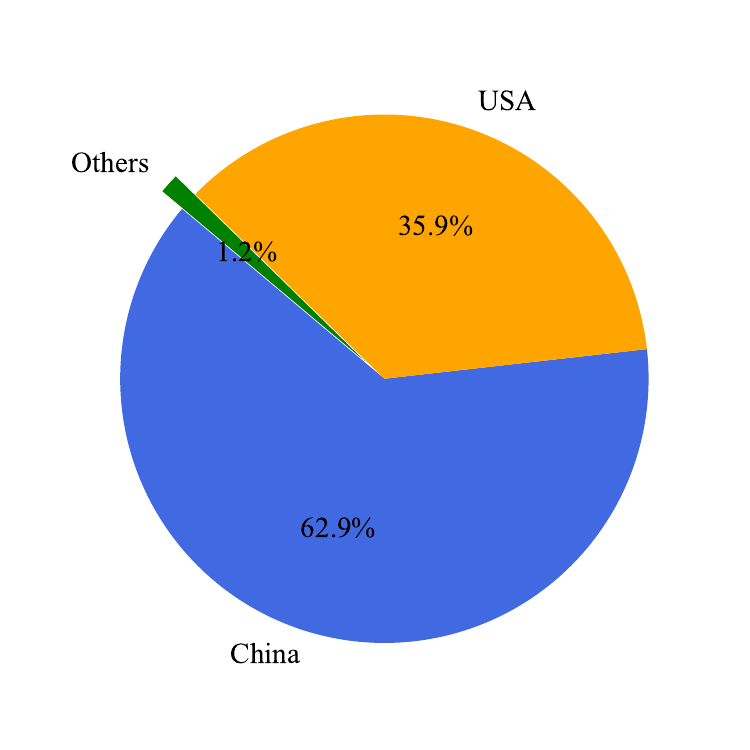}
    \caption{Geographical Distribution of Survey Respondents}
    \label{fig:geo_dist}
\end{figure}

\subsection{Geographical Distribution}
Figure \ref{fig:geo_dist} shows the geographical distribution of our previous survey. Among our 804 participants, 62.9\% of them are from China, 35.9\% from the USA, and 1.2\% from other regions. This highlights the dominance of responses from China and the USA.

\subsection{Results of the survey} \label{res_sur}

\begin{figure}[t]
    \centering
    \includegraphics[width=1\linewidth]{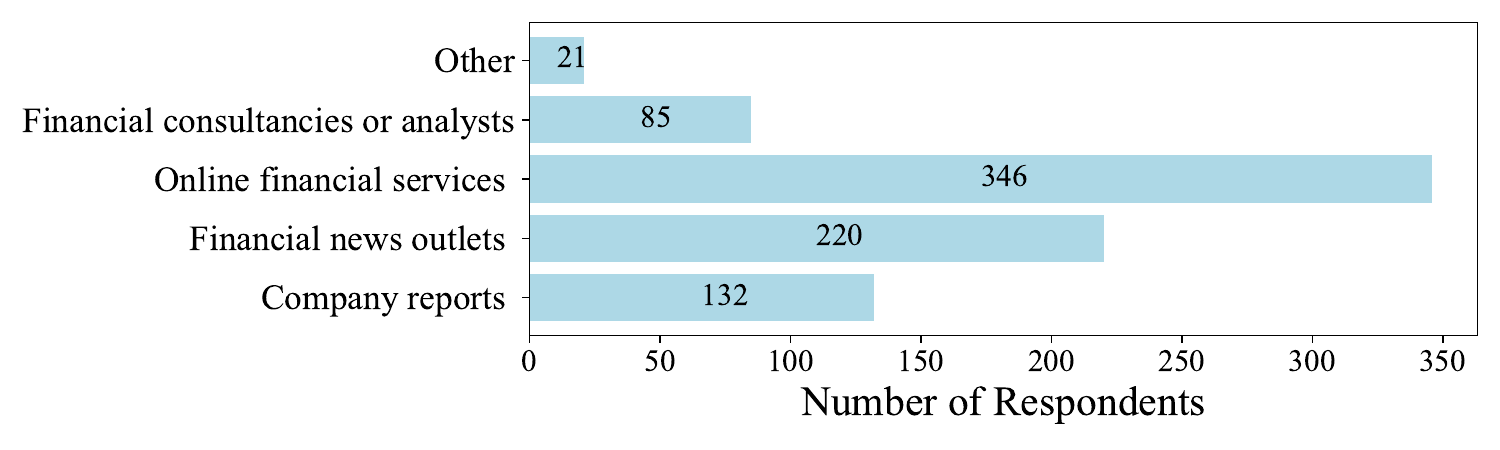}
    \caption{Primary Source of Financial Information extracted from the survey}
    \label{fig:survey_res1}
\end{figure}

\begin{figure}[t]
    \centering
    \includegraphics[width=1\linewidth]{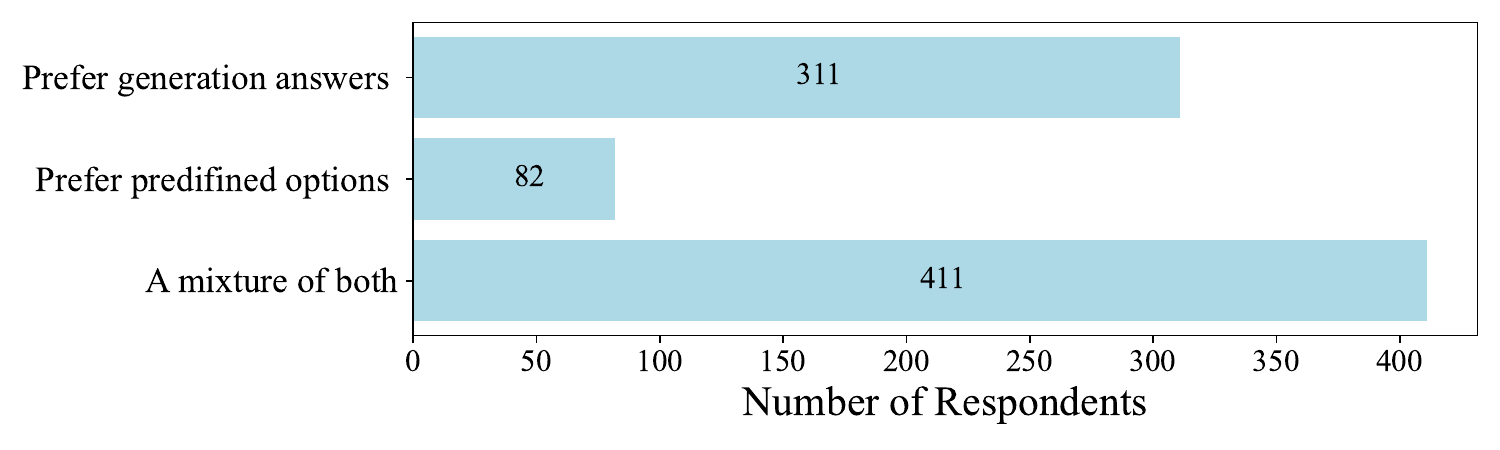}
    \caption{Results of whether preferring generation answers or predefined options}
    \label{fig:survey_res2}
\end{figure}

Figure \ref{fig:survey_res1} and Figure \ref{fig:survey_res2} report the primary financial information source and the results of whether users prefer generation answers or predefined options, which demonstrates the diversity of our benchmark contributors.

\subsection{Detailed Information of Each Task}

\subsubsection{Zero-shot Tasks}

Similar to existing benchmarks, the zero-shot tasks in the \textbf{UCFE benchmark} require models to handle new financial problems without any prior examples. These tasks assess the models' ability to generalize across different types of financial challenges. The \textbf{UCFE benchmark} includes four zero-shot tasks:
\begin{itemize}
    \item \textbf{Stock Price Prediction:} Predicting future stock prices using historical A-stock statistics is a common task in financial forecasting.
    \item \textbf{Bearish Information Detection:} Identifying whether the information affects the market negatively from sources such as EastMoney, similar to risk detection tasks in other benchmarks.
    \item \textbf{Financial Indicator Calculation:} Computing important financial metrics using standard CPA and CFA formulas, much like quantitative tasks in existing financial benchmarks.
    \item \textbf{Financial Information Summarization:} Summarizing news headlines to capture key insights, a task also present in general NLP benchmarks but adapted to the financial context.
\end{itemize}

These tasks are designed to reflect real-world financial decision-making scenarios, targeting two broad user groups: the general public and financial professionals. In essence, they encompass a wide range of users, making the benchmark applicable to all types of financial stakeholders.

\subsubsection{Few-shot Tasks}

The few-shot tasks in the \textbf{UCFE benchmark} involve multi-turn financial interactions, focusing on how models adapt to evolving user input over several rounds. Unlike single-turn tasks in existing benchmarks, these tasks emphasize real-world financial decision-making. We categorize the 13 tasks into the following four main groups:

\paragraph{Analytical and Evaluation Tasks}
These tasks require the model to simulate the role of financial analysts, providing detailed insights based on iterative queries. The model must refine its responses as users ask follow-up questions:
\begin{itemize}
    \item \textbf{Analyst Simulation:} Comprehensive analysis of company performances from financial reports and analyst reviews and generate recommendations.
    \item \textbf{Asset Valuation Reporting:} Provide asset valuations using data from EastMoney.
    \item \textbf{Company Evaluation Reporting:} Evaluate company performance using financial reports.
    \item \textbf{Corporate Operation Analysis:} Analyze company operations based on analyst reports.
\end{itemize}

\paragraph{Risk and Compliance Tasks}
These tasks focus on identifying financial risks and ensuring compliance with regulations, where users interact with the model to iteratively refine their analysis:
\begin{itemize}
    \item \textbf{Credit Risk Evaluation:} Assess credit risks based on GPT-4-generated data.
    \item \textbf{Insider Trading Detection:} Identify potential insider trading cases using court records of historical insider trading case reports.
    \item \textbf{Prospectus Risk Summarization:} Summarize risks in prospectuses and inquiry letters, refining insights based on user feedback.
    \item \textbf{Financial Regulation Consulting:} Provide guidance on regulatory compliance and potential punishments using Securities Law.
\end{itemize}

\paragraph{Strategy and Optimization Tasks}
In these tasks, users interact with the model to evaluate and optimize investment strategies. The multi-turn nature allows users to explore different strategies or fine-tune their approach:
\begin{itemize}
    \item \textbf{Investment Strategy Evaluation:} Evaluate effectiveness and summarize investment strategies using data from Seeking Alpha.
    \item \textbf{Investment Strategy Optimization:} Optimize strategies with feedback from multiple rounds of user queries.
\end{itemize}

\paragraph{Consulting and Summarization Tasks}
These tasks involve providing consulting services or summarizing financial information, where users may request additional clarification or focused insights over several interactions:
\begin{itemize}
    \item \textbf{Financial Knowledge Consulting:} Offer explanation on financial terminologies and basic financial knowledge based on sources like Investopedia.
    \item \textbf{Industry Report Summarization:} Summarize industry reports from EastMoney, allowing users to quickly identify key trading insights.
    \item \textbf{Newshare Evaluation Reporting:} Evaluate target price range of newly issued shares, analyze risk and opportunity of the new share based on company overview using data from platforms like \url{stock.us}.
\end{itemize}

\begin{figure}[htbp]
    \centering
    \includegraphics[width=1\linewidth]{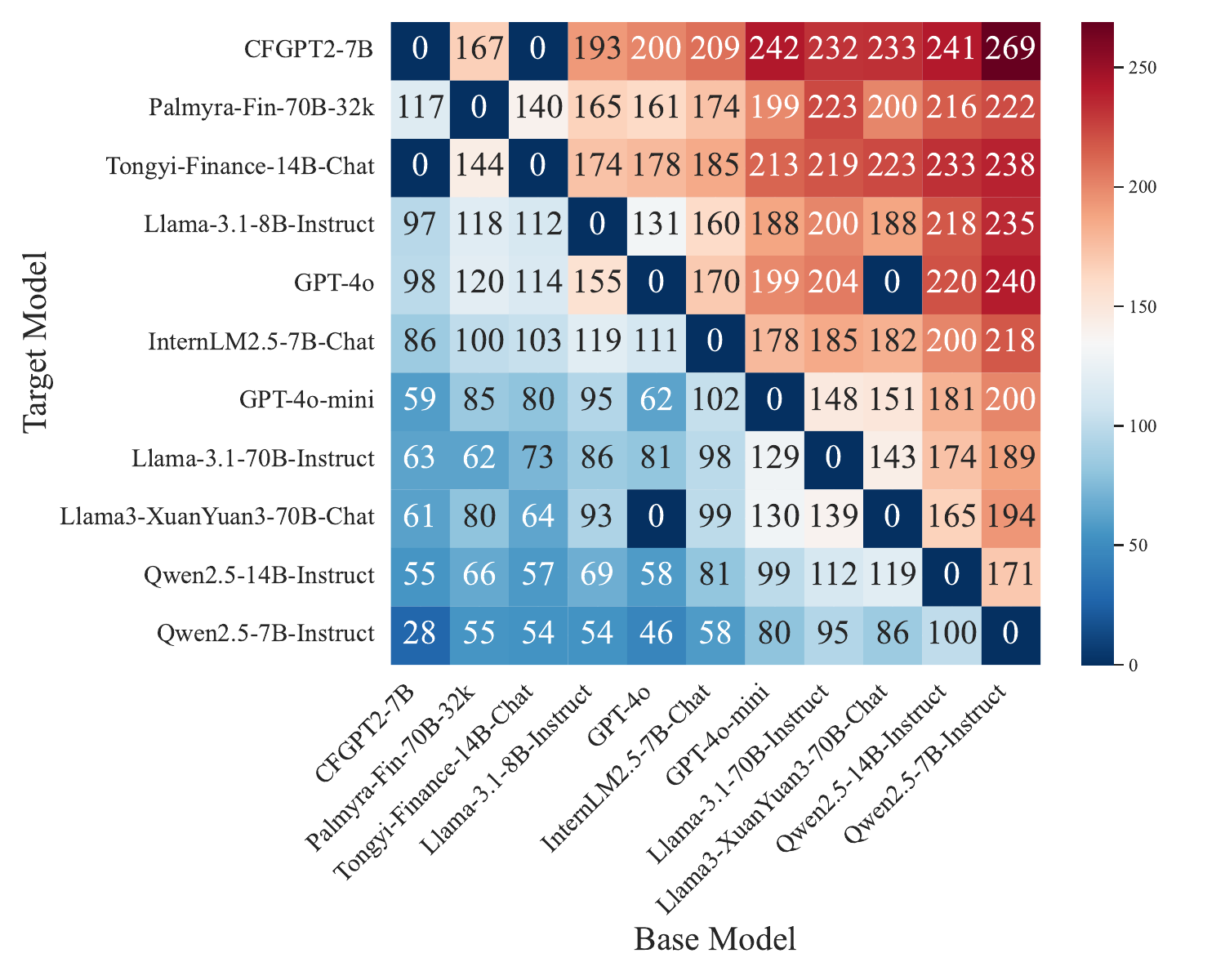}
\caption{Win counts heatmap for all tasks. The heatmap illustrates the total number of wins where the target model outperforms the base model across all head-to-head comparisons.}
    \label{fig:heatmap}
\end{figure}

\section{Human Expert Evaluation}

\begin{figure}[t]
    \centering
    \includegraphics[width=1\linewidth]{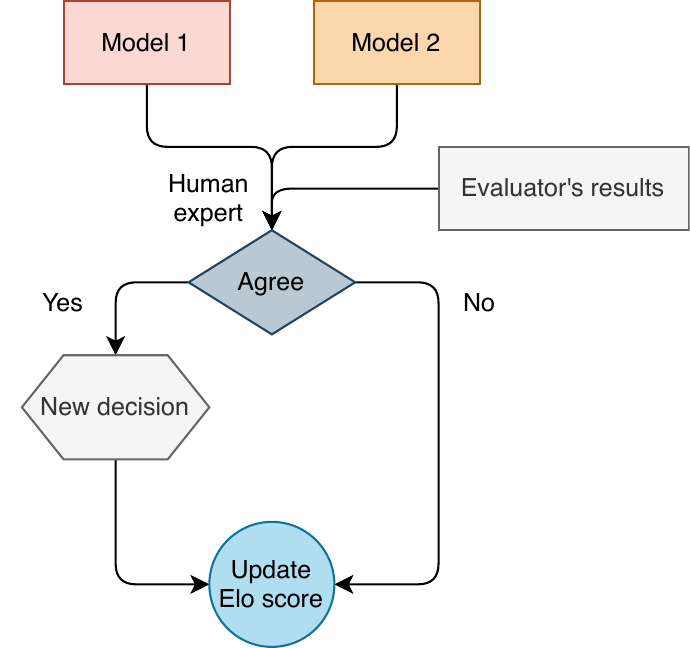}
    \caption{Human evaluation pipeline.}
    \label{fig:human_eval}
\end{figure}

\begin{figure*}[t]
    \centering
    \includegraphics[width=1.1\textwidth,height=0.9\textheight,keepaspectratio]{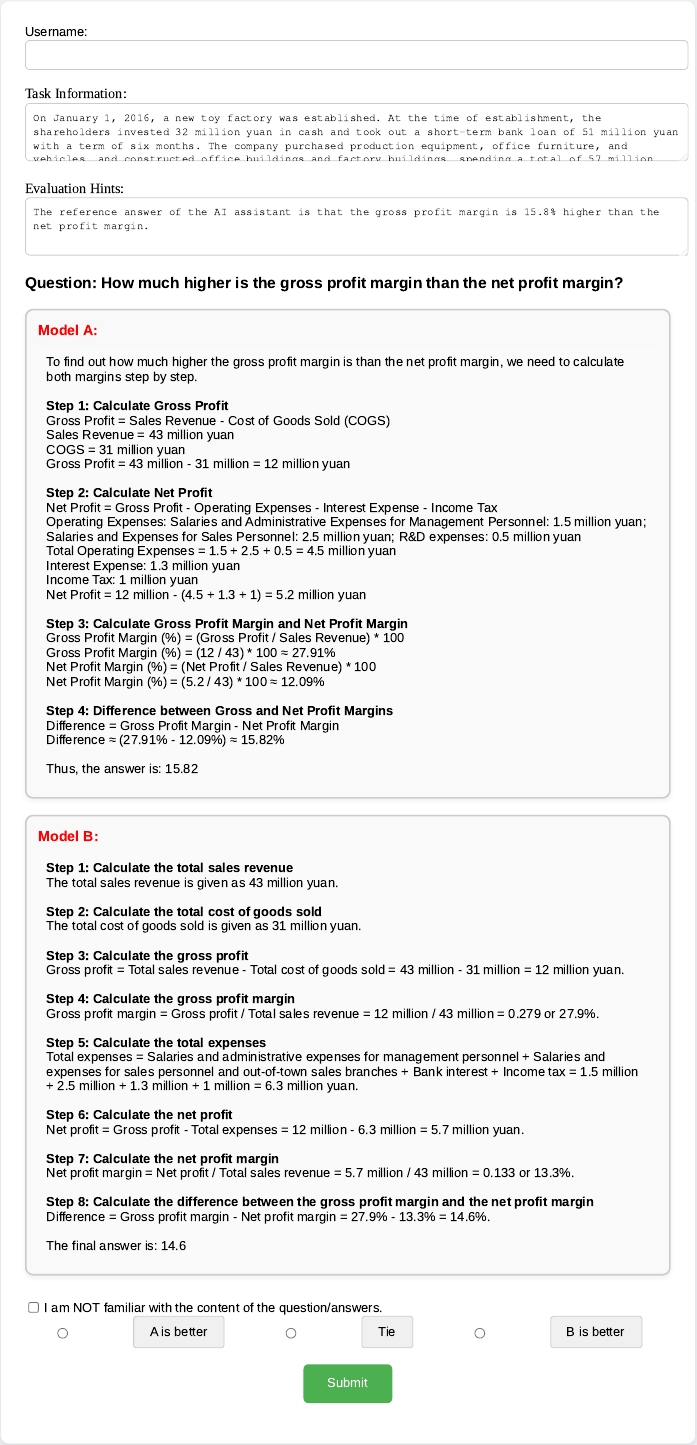}
    \caption{UI interface for human expert evaluation}
    \label{fig:ui}
\end{figure*}

Figure \ref{fig:human_eval} illustrates the evaluation pipeline. Human experts first assess the existing results, and if the outcomes directly contradict their expertise, they are asked to make another selection. Additionally, we designed a UI interface, shown in Figure \ref{fig:ui}, to ensure that participants are unaware of the model names. After all the evaluations, we recalculated the Elo scores based on the samples evaluated by the human experts.

\begin{figure*}[htbp]
\centering
\small
\begin{tcolorbox}[title = {Example Dialogue}, colframe=yellow!70!black, colback=yellow!10, coltitle=white, colbacktitle=yellow!50!black]
\textbf{User Simulator:}
Hey, can you explain what financial status analysis is? I'm curious about how it's used by businesses or individuals.\\
\textbf{\\LLM:}
Sure, I'd be happy to explain the financial status analysis! Financial status analysis is a process of evaluating an individual's or a business's financial situation to determine their current financial health and identify areas for improvement. It involves analyzing various financial statements and documents to assess their financial performance, liquidity, solvency, and overall financial position. For individuals, financial status analysis may include reviewing bank statements, credit reports, tax returns, and other financial documents to determine their income, expenses, assets, and liabilities. This information can help individuals understand their current financial situation, identify areas for improvement, and create a plan to achieve their financial goals. For businesses, financial status analysis may involve reviewing financial statements, such as the balance sheet, income statement, and cash flow statement, to assess their financial performance, liquidity, and solvency. This information can help businesses identify areas where they can improve their financial performance, make informed decisions about investments and spending, and manage their finances more effectively. Overall, financial status analysis is an important tool for individuals and businesses to understand their financial situation and make informed decisions about their finances.\\
\textbf{\\User Simulator:} Alright, that makes sense. So, for a business, does this analysis help in making decisions about spending and investments based on the financial statements?\\
\textbf{\\LLM:} Yes, that's correct! Financial status analysis for a business involves reviewing financial statements to assess the company's financial performance, liquidity, solvency, and overall financial position. This information can help the business make informed decisions about spending and investments. For example, if a business is experiencing a high level of liquidity, it may be able to invest more in new equipment or marketing initiatives to grow its business. On the other hand, if the business is experiencing financial difficulties, it may need to focus on improving its cash flow and reducing expenses to maintain solvency. By analyzing financial statements, a business can identify areas where it can improve its financial performance and make informed decisions about spending and investments. This can help the business achieve its long-term goals and maintain financial stability.\\
\textbf{\\User Simulator:} Got it, thanks! So it's all about understanding where the money's coming from and going, right?\\
\textbf{\\LLM:} Yes, that's right! Financial status analysis involves understanding where the money is coming from and going to. By analyzing financial statements and other financial documents, individuals and businesses can gain insight into their income, expenses, assets, and liabilities. For individuals, this can help them understand their cash flow and identify areas where they can reduce expenses or increase their income. For businesses, it can help them identify areas where they can improve their financial performance and make informed decisions about spending and investments. Overall, financial status analysis is an important tool for individuals and businesses to understand their financial situation and make informed decisions about their finances.\\
\textbf{\\User Simulator:} Great, that clears things up for me. Thanks for the help! <Consultation Ended>
\end{tcolorbox}
\caption{Example Dialogue between User Simulator and Tongyi-Finance-14B-Chat}
\label{example_dial}
\end{figure*}

\section{More Experiment Results}

Figure~\ref{fig:heatmap} highlights each model's comparative performance, showcasing strengths and weaknesses. And Table \ref{tab:all} presents the Elo scores across all tasks. The results show that LLMs trained on financial text data demonstrate outstanding performance across various tasks. However, due to limitations in the zero-shot task texts, the models' performance in those tasks may be affected.

Figure \ref{example_dial} shows an example dialogue between User Simulator and Tongyi-Finance in the ``Financial Knowledge Consulting'' Task.

\begin{sidewaystable}[t]
\centering
\renewcommand{\arraystretch}{1} 
\resizebox{1\textwidth}{!}{
\begin{threeparttable} 
\begin{tabular}{@{}l c c c c c c c c c c c @{}}
\toprule
\textbf{Model} & \textbf{Tongyi-Finance-14B-Chat} & \textbf{CFGPT2-7B} & \textbf{Palmyra-Fin-70B-32k} & \textbf{GPT-4o} & \textbf{Llama-3.1-8B-Instruct} & \textbf{Internlm2.5-7b-chat} & \textbf{Llama3-XuanYuan3-70B-Chat} & \textbf{Llama-3.1-70B-Instruct} & \textbf{GPT-4o-mini} & \textbf{Qwen2.5-14B-Instruct } & \textbf{Qwen2.5-7B-Instruct}\\
\midrule
\textbf{Analyst Simulation} & \colorbox{blue!10}{\textbf{1114.25}} & \colorbox{red!10}{\textbf{1188.80}} & 1036.15 & 1066.31 & 1027.62 & 1018.49 & 926.93 & 915.40 & 933.63 & 875.52 & 895.78 \\
\textbf{Asset Valuation Reporting} & 1077.59 & \colorbox{red!10}{\textbf{1202.57}} & \colorbox{blue!10}{\textbf{1142.88}} & 1033.36 & 959.37 & 1060.53 & 871.43 & 939.43 & 956.57 & 931.89 & 823.02 \\
\textbf{Company Evaluation Reporting} & \colorbox{blue!10}{\textbf{1068.18}} & \colorbox{red!10}{\textbf{1194.01}} & 1000.03 & 1038.43 & 1027.47 & 1047.91 & 886.98 & 917.06 & 1043.14 & 891.65 & 883.52 \\
\textbf{Corporate Operation Analysis} & \colorbox{red!10}{\textbf{1178.15}} & 1053.48 & \colorbox{blue!10}{\textbf{1097.44}} & 1076.33 & 1020.20 & 909.57 & 980.96 & 1006.09 & 977.73 & 909.79 & 788.69 \\
\textbf{Credit Risk Evaluation}& \colorbox{blue!10}{\textbf{1151.50}} & \colorbox{red!10}{\textbf{1209.96}} & 1121.60 & 1082.85 & 1028.88 & 1039.29 & 992.64 & 956.07 & 888.36 & 806.93 & 719.89 \\
\textbf{Financial Knowledge Consulting} & \colorbox{blue!10}{\textbf{1181.90}} & \colorbox{red!10}{\textbf{1275.16}} & 1130.83 & 1073.20 & 1062.81 & 958.14 & 920.05 & 902.51 & 875.90 & 869.82 & 747.07 \\
\textbf{Financial Regulation Consulting} & 1070.82 & \colorbox{red!10}{\textbf{1159.60}} & \colorbox{blue!10}{\textbf{1128.95}} & 1035.19 & 1034.53 & 1025.16 & 1008.68 & 982.39 & 891.72 & 865.44 & 795.66\\
\textbf{Industry Report Summarization}  & 996.86 & 1035.47 & \colorbox{red!10}{\textbf{1154.38}} & \colorbox{blue!10}{\textbf{1149.73}} & 1050.53 & 1003.94 & 888.71 & 928.58 & 978.90 & 947.57 & 863.84 \\
\textbf{Insider Trading Detection} & 1031.29 & 1036.04 & \colorbox{blue!10}{\textbf{1139.13}} & 1058.42 & \colorbox{red!10}{\textbf{1180.18}} & 1054.34 & 936.75 & 968.44 & 975.41 & 855.87 & 763.09 \\
\textbf{Investment Strategy Evaluation} & \colorbox{red!10}{\textbf{1223.95}} & \colorbox{blue!10}{\textbf{1173.37}} & 1074.47 & 999.73 & 1086.70 & 969.71 & 929.73 & 981.23 & 928.18 & 843.62 & 787.67 \\
\textbf{Investment Strategy Optimization} & \colorbox{red!10}{\textbf{1177.51}} & \colorbox{blue!10}{\textbf{1140.73}} & 1131.13 & 997.38 & 1087.51 & 1035.26 & 961.40 & 981.92 & 898.98 & 829.85 & 756.83 \\
\textbf{Newshare Evaluation Reporting} & 989.39 & 1037.57 & \colorbox{blue!10}{\textbf{1056.65}} & \colorbox{red!10}{\textbf{1114.74}} & 1030.47 & 982.26 & 925.07 & 879.8 & 1052.01 & 1050.42 & 880.39 \\
\textbf{Prospectus Risk Summarization} & \colorbox{red!10}{\textbf{1171.76}} & \colorbox{blue!10}{\textbf{1123.44}} & 1122.57 & 1110.57 & 979.56 & 953.88 & 899.72 & 941.71 & 922.92 & 835.63 & 936.02 \\
\midrule
\textbf{Stock Price Prediction} & \colorbox{blue!10}{\textbf{1018.66}} & 981.21 & 983.73 & 1012.44 & 967.88 & 1003.67 & 985.76 & 1000.01 & \colorbox{red!10}{\textbf{1024.75}} & 1011.70 & 1010.08 \\
\textbf{Negative Information Detection} & 998.72 & \colorbox{red!10}{\textbf{1113.40}} & 998.85 & 978.35 & 1007.33 & 978.85 & 970.53 & 980.32 & 980.59 & \colorbox{blue!10}{\textbf{1009.53}} & 983.24 \\
\textbf{Financial Indicator Calculation} & 1119.17 & \colorbox{red!10}{\textbf{1180.77}} & 1067.67 & 858.10 & \colorbox{blue!10}{\textbf{1131.91}} & 1061.85 & 972.09 & 928.13 & 861.15 & 897.84 & 919.85 \\
\textbf{Financial Text Summarization} & 974.17 &  \colorbox{red!10}{\textbf{1103.51}} & 1030.04 & 1022.80 & \colorbox{blue!10}{\textbf{1095.63}} & 982.84 & 909.35 & 1027.29 & 954.68 & 971.29 & 927.49 \\

\midrule
\textbf{Overall} & \colorbox{red!10}{\textbf{1156.99}}
& \colorbox{blue!10}{\textbf{1155.75}} & 1128.25 & 1117.68 & 1046.87 & 995.85 & 913.48 & 912.26 & 901.75 & 855.82 & 814.48\\
\bottomrule
\end{tabular}
\end{threeparttable}
}
\caption{Model results in all the tasks of the \textbf{UCFE benchmark}. \colorbox{red!10}{Red} highlights the highest value, while \colorbox{blue!10}{Blue} represents the second-highest value.}
\label{tab:all}
\end{sidewaystable}

\section{Prompt} \label{prompt}

\subsection{Zero Shot Task}

Figures [\ref{zero_test}, \ref{zero_source}, \ref{zero_eval}] have shown all the prompts we used for testing and evaluation for \textit{Financial Text Summarization}.

\begin{figure*}[htbp]
\centering
\small

\begin{tcolorbox}[title = {Test Prompt}, colframe=gray, colback=gray!10, coltitle=white, colbacktitle=gray!50!black]
\textbf{Model Prompt:\\}
You are providing a summary service for financial texts to help users extract key points from complex financial information.\\The given financial text is: \{information\}\\Your task is: \{needs\}.
\end{tcolorbox}
\caption{Test Prompt for \textit{Financial Text Summarization}}
\label{zero_test}
\end{figure*}

\begin{figure*}[htbp]
\centering
\small
\begin{tcolorbox}[title = {Source Information}, colframe=blue, colback=blue!10, coltitle=white, colbacktitle=blue!50!black]
New Zealand's Ministry of Foreign Affairs issued a statement on the 19th regarding the volcanic eruption disaster in Tonga, stating that Tonga has now established a temporary communication system that can use 2G signals to contact the outside world, but communication is still ``limited and sporadic.'' Meanwhile, countries such as New Zealand and Fiji plan to provide aid to Tonga.
\end{tcolorbox}
\caption{Source Information for \textit{Financial Text Summarization}}
\label{zero_source}
\end{figure*}

\begin{figure*}[htbp]
\centering
\small
\begin{tcolorbox}[title = {Evaluation Prompt}, colframe=yellow!70!black, colback=yellow!10, coltitle=white, colbacktitle=yellow!50!black]
\textbf{Evaluation Criteria:\\}
Please act as a fair judge to assess the quality of the dialogue between the user and the AI assistant. Please read the user requirements and evaluation hints before assessing to help you better analyze the dialogue quality.\\ The user's needs are: \{needs\}.\\ The evaluation hints regarding specific content for your reference are: \{evaluation\_hints\} \\ When assessing, you also need to consider the following dimensions:\\ - Meeting user needs: Your evaluation should consider whether the AI assistant's responses comprehensively and appropriately meet the user's needs.\\- Accuracy of facts: Is the information provided accurate and based on credible facts and data?\\- Fairness and accountability: Are the suggestions or information provided feasible and accountable, and do they consider potential risks and consequences?\\- Richness: Does it contain abundant information, depth, contextual considerations, diversity, detailed explanations, and examples to meet user needs and provide comprehensive understanding?\\- Hallucination: Are there any hallucinations in the AI assistant's responses?\\- Note: Do not let the length of the response affect your scoring! Longer responses are not necessarily better; concise answers that meet the above requirements are good.\\ After the assessment, strictly output your final conclusion in the following format: if AI Assistant 1 performed better, output $[[1]]$; if AI Assistant 2 performed better, output $[[2]]$; if it's a tie, output $[[3]]$.\\\\$[$AI Assistant 1 Dialogue Start$]$\\\{dialogue1\}\\$[$AI Assistant 1 Dialogue End$]$\\\\$[$AI Assistant 2 Dialogue Start$]$\\\{dialogue2\}\\$[$AI Assistant 2 Dialogue End$]$\\\\
\textbf{Evaluation Points:\\}
AI assistant's suggested reference answer is: ``The New Zealand Ministry of Foreign Affairs stated on the 19th that Tonga has established a temporary communication system, but communication is still limited. New Zealand and countries like Fiji plan to provide aid to Tonga.'' This answer is an ideal response example.\\The AI assistant's response must include the following key content (the expression can vary): ``Tonga has established a temporary communication system,'' and the absence of this content will result in the answer being directly judged as incorrect.\\Ideally, the AI assistant's response should include the following points: ``communication is limited, New Zealand and countries like Fiji plan to provide aid to Tonga,'' to ensure the comprehensiveness of the answer.
\end{tcolorbox}
\caption{Evaluation Prompt for \textit{Financial Text Summarization}}
\label{zero_eval}
\end{figure*}

\subsection{Few Shot Task}

Figures [\ref{few_test}, \ref{few_source}, \ref{few_eval}] have shown all the prompts we used for testing and evaluation for \textit{Asset Valuation Reporting}.

\begin{figure*}[htbp]
\centering
\small

\begin{tcolorbox}[title = {Test Prompt}, colframe=gray, colback=gray!10, coltitle=white, colbacktitle=gray!50!black]
\textbf{Role Prompt:\\}
You are role-playing as a writer.\\ You are conversing with an AI assistant, hoping it can help generate an asset evaluation report.\\ The purpose, object, and scope of the evaluation are: {information}. Your needs are: {needs}.\\ Ensure to converse with the AI assistant in the tone of a writer, avoid unnecessary chatter, and try to be as realistic as possible.\\ If you feel the AI assistant's response meets your needs, you can output the corresponding characters as instructed by the prompt. If not, raise your concerns based on the AI assistant's response.\\ Note: What you need to do is simulate a user asking the AI assistant questions based on the provided information and needs (if any) rather than answering or solving problems.\\ You do not need to perform any calculations, analysis, or generate report content. If the AI assistant asks questions or needs additional information, please answer truthfully.\\ Please start your conversation.
\\
\\
\textbf{User Intention:\\}
Generate the purpose, object, and scope sections in the asset evaluation report.\\
\\
\textbf{Model Prompt:\\}
You are providing document services to a writer. During the service provision process, you can ask the other party for more information. The template must be in the form of an asset appraisal report.
\end{tcolorbox}
\caption{Test Prompt for \textit{Asset Valuation Reporting}}
\label{few_test}
\end{figure*}

\begin{figure*}[htbp]
\centering
\small
\begin{tcolorbox}[title = {Source Information}, colframe=blue, colback=blue!10, coltitle=white, colbacktitle=blue!50!black]
The evaluation object is Keda Guochuang Xinneng Technology Co., Ltd., and the reason for evaluating the assets is that Keda Guochuang Software Co., Ltd. is issuing shares to purchase 100\% equity of Keda Guochuang Xinneng Technology Co., Ltd. Keda Guochuang Software Co., Ltd. has signed the ``Share Issuance and Asset Purchase Agreement'' with Hefei Guibo Equity Investment Partnership (Limited Partnership) and Hefei Zixu Investment Partnership (Limited Partnership). The partners of Hefei Guibo Equity Investment Partnership (Limited Partnership) and Hefei Zixu Investment Partnership (Limited Partnership) include Sun Lu, Shi Xingling, Xu Genyi, Chen Xuexiang, Zhang Qiyun, and Dong Xianquan. As of December 31, 2020, the book value of the company's total assets was 461.3236 million yuan, the book value of total liabilities was 161.9956 million yuan, and the book value of net assets was 299.3280 million yuan.
\end{tcolorbox}
\caption{Source Information for \textit{Asset Valuation Reporting}}
\label{few_source}
\end{figure*}

\begin{figure*}[htbp]
\centering
\small
\begin{tcolorbox}[title = {Evaluation Prompt}, colframe=yellow!70!black, colback=yellow!10, coltitle=white, colbacktitle=yellow!50!black]
\textbf{Evaluation Criteria:\\}
Please act as a fair judge to assess the quality of the dialogue between the user and the AI assistant. Please read the user requirements and evaluation hints before assessing to help you better analyze the dialogue quality.\\ The user's needs are: \{needs\}.\\ The evaluation hints regarding specific content for your reference are: \{evaluation\_hints\} \\ When assessing, you also need to consider the following dimensions:\\ - Meeting user needs: Your evaluation should consider whether the AI assistant's responses comprehensively and appropriately meet the user's needs.\\- Accuracy of facts: Is the information provided accurate and based on credible facts and data?\\- Fairness and accountability: Are the suggestions or information provided feasible and accountable, and do they consider potential risks and consequences?\\- Richness: Does it contain abundant information, depth, contextual considerations, diversity, detailed explanations, and examples to meet user needs and provide comprehensive understanding?\\- Hallucination: Are there any hallucinations in the AI assistant's responses?\\- Note: Do not let the length of the response affect your scoring! Longer responses are not necessarily better; concise answers that meet the above requirements are good.\\ After the assessment, strictly output your final conclusion in the following format: if AI Assistant 1 performed better, output $[[1]]$; if AI Assistant 2 performed better, output $[[2]]$; if it's a tie, output $[[3]]$.\\\\$[$AI Assistant 1 Dialogue Start$]$\\\{dialogue1\}\\$[$AI Assistant 1 Dialogue End$]$\\\\$[$AI Assistant 2 Dialogue Start$]$\\\{dialogue2\}\\$[$AI Assistant 2 Dialogue End$]$\\\\
\textbf{Evaluation Points:\\}
AI assistant's response must include the following key content (expression can vary): 1. The assessment purpose must include the agreement signing time; 2. The partner's name must be included in the assessment purpose section; 3. The assessment purpose must be a single section; 4. The assessment object and scope must be the second section, and content from different sections should not be confused, missing these contents will result in the answer being directly judged as incorrect.\\Encourage the AI assistant to mention the following content in the response: 1. Add background information to enhance the completeness of the report. 2. Provide a detailed explanation of the assessment purpose, including an explanation of the importance or necessity of the assessment purpose, to improve the persuasiveness of the report. 3. Any key factors within the assessment scope that may affect asset value, such as market conditions, industry trends, etc., to provide a more comprehensive assessment perspective, this will help improve the quality of the answer.\\The AI assistant's response should avoid including the following content: 1. Avoid including detailed company history or unrelated business introductions that are not related to the assessment purpose, object, and scope in the report. 2. Vague or uncertain language. 3. Ensure that the assessment purpose, object, and scope are each independent, do not mix the content together. Mentioning these contents will result in the answer being judged as inappropriate.
\end{tcolorbox}
\caption{Evaluation Prompt for \textit{Asset Valuation Reporting}}
\label{few_eval}
\end{figure*}

\end{document}